%% file: main.tex
\DocumentMetadata{}
\documentclass[manuscript,screen,nonacm]{acmart}

\usepackage{hyperref}
\usepackage{hyperxmp}
\usepackage{amsmath,amssymb,amsfonts}
\usepackage{graphicx}
\usepackage{textcomp}
\usepackage{makecell}
\usepackage{bbding}
\usepackage{pifont}
\usepackage{soul,xcolor}
\usepackage{tcolorbox}
\usepackage{colortbl}
\usepackage{caption}
\usepackage{subcaption}
\usepackage{multirow}
\usepackage{booktabs}
\usepackage{tabularx}
\usepackage{array}
\usepackage{algorithm} % for pseudocode float
\usepackage{algpseudocode} % for pseudocode
\usepackage{amssymb}
\usepackage{amsmath}
\usepackage{cellspace} % 新添加的包
\usepackage{xspace}
\usepackage{paralist}
\usepackage{tikz}
\usepackage{ulem}
\AtBeginDocument{%
  }

\setcopyright{acmlicensed}
\copyrightyear{2024}
\acmYear{2024}
\acmDOI{XXXXXXX.XXXXXXX}
\acmISBN{978-1-4503-XXXX-X/18/06}
\makeatletter
\DeclareRobustCommand\onedot{\futurelet\@let@token\@onedot}
\def\@onedot{\ifx\@let@token.\else.\null\fi\xspace}
\def\eg{\emph{e.g}\onedot} 
\def\ie{\emph{i.e}\onedot}

\makeatother

\begin{document}

\title{Antidote or Placebo? Unraveling the Efficacy of Neuron Coverage Criteria on Testing Transformer-based Language Models}

%\author{Ben Trovato}
%\authornote{Both authors contributed equally to this research.}
%\email{trovato@corporation.com}
%\orcid{1234-5678-9012}
%\author{G.K.M. Tobin}
%\authornotemark[1]
%\email{webmaster@marysville-ohio.com}
%\affiliation{%
%  \institution{Institute for Clarity in Documentation}
%  \streetaddress{P.O. Box 1212}
%  \city{Dublin}
%  \state{Ohio}
%  \country{USA}
%  \postcode{43017-6221}
%}
%
\author{Xiaoning Ren}
\affiliation{%
  \institution{University of Science and Technology of China}
  \country{China}}
\email{hnurxn@mail.ustc.edu.cn}

\author{Jiayang Song}
\affiliation{%
	\institution{Macau University of Science and Technology}
	\country{China}}
\email{jiayang.song@ieee.org}

\author{Chongyang Liu}
\affiliation{%
	\institution{University of Science and Technology of China}
	\country{China}}
\email{lcyyy@mail.ustc.edu.cn}

\author{Jie Li}
\affiliation{%
	\institution{University of Science and Technology of China}
	\country{China}}
\email{lijie2000@mail.ustc.edu.cn}

\author{Lei Ma}
\affiliation{%
	\institution{The University of Tokyo}
	\country{Japan}}
\affiliation{%
	\institution{University of Alberta}
	\country{Canada}}
\email{ma.lei@acm.org}

\author{Yinxing Xue}
\authornote{Corresponding author.}
\affiliation{%
	\institution{Chinese Academy of Sciences, Institute of AI for Industries}
	\country{China}}
\email{yxxue@iaii.ac.cn}

%
%\author{Valerie B\'eranger}
%\affiliation{%
%  \institution{Inria Paris-Rocquencourt}
%  \city{Rocquencourt}
%  \country{France}
%}

%%
%% By default, the full list of authors will be used in the page
%% headers. Often, this list is too long, and will overlap
%% other information printed in the page headers. This command allows
%% the author to define a more concise list
%% of authors' names for this purpose.
% \renewcommand{\shortauthors}{Trovato et al.}

%%
%% The abstract is a short summary of the work to be presented in the
%% article.

\begin{abstract}
  In the realm of deep learning, a variety of neuron coverage criteria for Deep Neural Networks (DNNs)  have been devised to effectively assess the quality of test suites and facilitate the generation of test inputs. 
  Recently proposed coverage criteria, incorporating representation distribution and causal relationships, have infused fresh vitality into this field.
  However, the focus of previous works is primarily on Convolutional Neural Networks for computer vision, leading to a research gap in exploring coverage testing for language models.
  Concurrently, with the rise of large language models, transformer-based language models have become increasingly dominant, and numerous ones have sprouted.
  Therefore, the effectiveness of coverage criteria in transformer-based language tasks, especially with the introduction of novel criteria, remains an unresolved open problem. 
  To tackle it, this study examines these concerns by evaluating a wide range of criteria, including four well-established ones and two state-of-the-art criteria, across three types of transformer-based models: encoder-only, decoder-only, and encoder-decoder models.
  
  Building on previous research, we conduct a comprehensive evaluation across three key areas:
  regarding test suite properties, 1) Error-revealing capability, i.e., sensitivity to adversarial examples;
  2) Diversity, i.e., distribution diversity and sample fairness (category diversity); 
  and regarding test suite generation, 3) Input generation guidance, i.e., the ability to guide the generation of more valuable samples.
  The experimental results demonstrate that the impact of coverage criteria is multifaceted. For the error-revealing capability of test suites, the additional coverage for erroneous samples over noise samples is only 0.32\%. In terms of distribution diversity and sample fairness, 26 and 30 cases, respectively, out of 33 configurations are effectively evaluated. Additionally, incorporating neuron-wise coverage guidance during test suite generation slightly increases the production of adversarial samples by 4.56\%. 
  
  In conclusion, while current coverage criteria can act as an antidote for assessing simple diversity, they remain largely a placebo for the core task of revealing adversarial errors, particularly when relying on individual criterion.
  Consequently, their practical application requires carefully evaluating the trade-off between computational overhead and potential benefits given the massive scale of Transformers.
  However, this low cost-effectiveness ultimately highlights the urgent need to explore and develop more robust and efficient criteria designed specifically for Transformer-based models.
  
\end{abstract}


	\maketitle

\input{chapter/intro}

\input{chapter/background}

\input{chapter/criteria}
\input{chapter/empirical}
\input{chapter/evaluation1}

\input{chapter/evaluation2}
\input{chapter/evaluation3}

\input{chapter/evaluation4}
\input{chapter/relatedwork}

\input{chapter/discussion}
\input{chapter/conlusion}

\section*{Acknowledgments}

This work was supported in part by the National Natural Science Foundation of China (Grant No. 61972373). It was also supported in part by JST CRONOS Grant (No. JPMJCS24K8), JSPS KAKENHI Grants (No. JP21H04877, No. JP23H03372, and No. JP24K02920), the Canada CIFAR AI Chairs Program, the Natural Sciences and Engineering Research Council of Canada, and the Autoware Foundation. In addition, Jiayang Song was supported by the MUST Faculty Research Grant (FRG).
\newpage
\bibliographystyle{ACM-Reference-Format}
\bibliography{ref}

\end{document}

%% file: chapter/intro.tex
\section{Introduction}
\label{sec:introduction}

The meteoric evolution of Deep Neural Networks (DNNs) has enabled and boosted the emergence of AI-driven systems across a diverse spectrum of domains, such as robotics, autonomous driving and medical care~\cite{badar2020application,lauriola2022introduction}.
Despite the promising and multifarious functionalities empowered by DNNs, akin to traditional software systems, DNNs also exhibit numerous underlying issues, such as the susceptibleness to bugs, degraded performance when dealing with cases outside the training data, and vulnerabilities against adversarial attacks.
Considering DNNs are widely deployed on a series of safety-critical applications (\eg, medical image analysis ~\cite{nair2020exploring,loftus2022uncertainty,egger2022medical} and autonomous vehicle control~\cite{loquercio2020general,choi2019gaussian}), software engineering (SE) researchers and industrial practitioners are devoted to preventing such unexpected behaviors with all their might. 
In particular, testing and evaluating the performance of DNNs from various perspectives have become crucial topics in the SE community to safeguard the quality and further advance the deployment of DNNs in more vital domains~\cite{wang2024survey}.

Recent research has shown that the decision logic of DNNs is sensitive to input biases~\cite{geirhos2018imagenet, brendel2019approximating}; namely, the comprehensiveness of the testing and evaluation of DNNs are associated with the representativeness of the test suites.
Therefore, testing the DNNs with complex characteristics necessitates effective test criteria to measure the degree of representativeness of the test suites and further guide the testing processing. 
In this context, a series of neuron coverage criteria have been proposed by software engineering practitioners~\cite{yan2024revisiting}. These criteria are primarily developed to evaluate the extent to which a test suite explores the behavior of a DNN by observing the outputs of neurons within the model. The more thoroughly the test suite explores the model’s behavior, the higher its representativeness. For instance, neuron coverage (NC)~\cite{deepxplore} quantifies the proportion of activated neurons during testing, which considers a higher neuron coverage implies a more comprehensive exploration of the behavioural space of the DNN model as well as an extensive understanding of the corresponding decision-making capabilities.
As a result, these criteria have become essential for assessing the quality of test suites and facilitating the effective testing of DNNs.

Interest in coverage testing techniques for DNNs has been steadily increasing. Since the introduction of DeepXplore by Pei et al. in 2017~\cite{deepxplore}, which pioneered the use of neuron coverage as a testing criterion, subsequent research~\cite{deepcover, deepcruiser, deepct, deepgauge, deeppath, ji2023cc, nlc} has expanded on this approach by developing novel criteria to improve DNN testing across various dimensions.
It is worth noting that most of these criteria are designed from the perspective of structural coverage; namely, the output of the structural elements (\ie, individual neurons or neuron combinations ) are utilized with specific operations to derive a statistical analysis regarding the degree of coverage of such elements.
In tandem, some critical analysis~\cite{harel2020neuron,dong2019there,li2019structural,yan2020correlations,yan2024revisiting,yang2022revisiting} of these criteria has emerged, highlighting limitations such as the insufficient correlation between some criteria and model quality, which pose concerns regarding the effectiveness and generalizability of these criteria. 
Despite these limitations, the positive impacts of coverage criteria have also been acknowledged~\cite{yan2024revisiting}. 
These discussions have catalyzed the ongoing evolution of coverage testing. 
Recent two studies, further encouraging and promoting the development of this field, demonstrated the efficacy of their coverage criteria~\cite{ji2023cc,nlc}, adopting representation distribution and causality, respectively.
Therefore, coverage testing remains crucial for assuring the quality of neural networks.

Unlike traditional DNNs, Large Language Models (LLMs) with transformer architecture~\cite{vaswani2017attention} as backbones are characterized by their self-attention mechanism, free-form auto-regressive generation format and generalization abilities across a diverse spectrum of tasks, which initiate a significant breakthrough in the field of Artificial Intelligence (AI). 
This innovation has recently gained substantial popularity due to the surge of LLM-empowered applications~\cite{bowman2015large,chowdhery2023palm,touvron2023llama}. 
With an increasing number of applications employing transformer-based models to extend their functionalities, it becomes even more vital to safeguard the quality and reliability of such models.
Nevertheless, previous work has primarily focused on fully connected networks and convolutional neural networks in the computer vision domain, with few attempts on transformer-based models. 
These models have a much larger number of neurons, increasing algorithmic complexity and leading to significantly higher computational costs.
Therefore, one question thereupon arises that:

\textbf{\textit{What and to what extent neuron coverage criteria are still effective in the context of testing transformer-based language models?}}

To explore this research question, in this paper, we are dedicated to investigating the effectiveness of a series of commonly studied neuron coverage criteria on different transformer-based language models. 
In particular, our first step involves a detailed review of existing neuron coverage criteria from two key perspectives: coverage granularity and algorithmic complexity. We categorize the neuron coverage criteria into three levels of granularity: single neuron, intra-layer neuron combinations, and inter-layer neuron combinations. Through a complexity analysis, we identify the criteria that can be applied to transformer-based language models, which have a large number of parameters while maintaining reasonable computational overhead. From this analysis, we select a representative set of criteria, which includes coverage criteria at all levels of granularity and incorporates two state-of-the-art criteria that have not been considered in earlier studies.
In the second step, we conduct a comprehensive evaluation of these criteria on transformer-based language models, focusing on the following aspects:
\textbf{1)Error-Revealing Capability} (RQ1): whether the criterion is sensitive to corner-case samples, thereby revealing the error-revealing capability of a test suite.
\textbf{2)Diversity} (RQ2 and RQ3): whether the criterion correlates positively with the distribution diversity and category diversity\footnote{Category diversity is also referred to as sample fairness. Throughout this paper, we use these two terms interchangeably.} of samples, reflecting the overall diversity of a test suite.
 \textbf{3)Input Generation Guidance}  (RQ4): whether the criterion can serve as effective guidance for selecting operators during the sample generation process, thereby aiding in the creation of valuable samples.

Extensive experiments are conducted, including seven coverage criteria and five transformer-based models on four downstream tasks.
The experimental results indicate that traditional neuron coverage criteria can be helpful for testing transformer-based language models in some cases, but in others, they offer little benefit. Specifically, for the test suite properties of distribution diversity and sample fairness, 26 and 30 cases, respectively, out of 33 configurations, are satisfied, indicating that most cases meet the criteria but with some exceptions. 
During the test suite generation process, adding neuron-wise coverage guidance slightly increases the generation of adversarial samples by 3.35\%. However, for error-revealing capability, the additional coverage for erroneous samples compared to noisy samples is only 0.32\%, suggesting that the coverage information gained from erroneous samples largely comes from the fact that they are novel inputs. Notably, coverage criteria such as NLC+, which are sensitive to the input order of the test suite, can exhibit a certain degree of randomness. 
Overall, when properly applied, coverage criteria can serve as an antidote, but if misapplied, they risk becoming a placebo. In other words, coverage criteria should be used as complementary tools, and testing can be enhanced by combining multiple criteria for more comprehensive evaluation.
Our contributions are outlined as follows:

\begin{compactitem}[$\bullet$]
	\item To the best of our knowledge, this study is the first to explore the efficacy of traditional neuron coverage criteria on transformer-based language models. Additionally, we consider a wide range of criteria, including two state-of-the-art criteria that have not been examined in earlier studies.
	\item We provide a novel analysis of most commonly applied existing neuron coverage criteria from the perspectives of coverage granularity and computational complexity, offering valuable insights for practitioners.
	\item We introduce a comprehensive evaluation covering three critical aspects: the error-revealing capability and diversity of test suite properties, as well as the input generation guidance during test suite creation.
	\item We replicate multiple traditional neuron coverage criteria on transformer-based models across various language tasks, and we open-source our implementation to facilitate the broader adoption and validation of transformer-based language models.
\end{compactitem}

To promote research on the safety of transformer-based language models in complex multilingual contexts, we make all source code publicly available at ~\href{https://github.com/hnurxn/LLMCoverageKit}{LLMCoverageKit}.
We provide the essential background and highlight the motivation in Section~\ref{sec:background}.
Section~\ref{sec:review} provides a comprehensive analysis of the target coverage.
Section~\ref{sec:study_design} details the workflow of our study and reports the corresponding experimental results.
Finally, Section~\ref{sec:related_work} covers related research work, Section~\ref{sec:discussion} discusses threats to validity,
and Section~\ref{sec:conclusion} concludes the paper.

%% file: chapter/background.tex
\section{Background and Motivation}
\label{sec:background}

\input{figure/figure2-1}

In the following section, we first provide essential information for neuron coverage criteria and transformer-based language models, then we introduce the current research gaps, summarize the open problems, and present the overall workflow of our study.
\subsection{Neuron Coverage Criteria}\label{section2-1}

Neuron coverage criteria are crucial techniques used in neural network testing to guide a comprehensive testing process. Following the simplest form of criterion NC by DeepXplore, several extended criteria have emerged. 
These criteria consider a single neuron or a set of associated neurons as an individual unit. 
Theoretically, neuron coverage criteria quantify the extent to which the constituent units of a neural network are explored by a test suite.
While the specific definition may vary based on context and analysis goals, a general definition of neuron coverage is as follows:

\emph{Definition 2.1.1 (Neuron Coverage)} Consider a neural network represented as $f$. 
It is common practice to abstract the network into multiple coverage units (based on one or more neurons), often defined by specific coverage criteria. 
For instance, we partition the neural network $f$ into $n$ coverage units, and this set of units can be denoted as $U=\{\text{unit}_{1}, \text{unit}_{2}, \ldots, \text{unit}_{n}\}$.  Let $\mathcal{D}$ be a test suite, where $f(\text{unit}_{k},\mathcal{D})$ denotes the coverage feature value of $\text{unit}_k$ when test suite $\mathcal{D}$ is input into the neural network, represented as $\mathcal{H}$.  Applying the feature values $\mathcal{H}$ to the coverage algorithm $\varphi$ results in $\varphi(\mathcal{H})$, which quantifies the coverage of $\text{unit}_{k}$. This can be expressed as follows:
\[\text{coverage}(f,\mathcal{D})=\sum_{\text{unit} \in U} \varphi(f(\text{unit},\mathcal{D})) / \| U \| .\]

\vspace{-2pt}

Intuitively, \emph{Definition 2.1.1} defines that for a given neural network $f$ and a test suite $\mathcal{D}$, each coverage criterion encompasses a unit partitioning method and a corresponding algorithm $\varphi$. The summation of coverage feature values across all units within the set $U$ yields the coverage value for the test suite when applied to the neural network $f$. 
At times, it has been attempted to compute the normalized value by dividing it by the total number of units $\| U \|$.
As shown in Figure ~\ref{figure2-1}, a neural network with multiple layers can be partitioned into various granularity units based on specific coverage criteria. For example, $\text{unit}_{1}$ represents a single neuron granularity, $\text{unit}_{2}$ represents a combination of neurons within the same layer, and $\text{unit}_{3}$ represents a combination of neurons across different layers. 
% Taking two specific neuron coverage criteria as examples.
Considering a case with two specific neuron coverage criteria, where a coverage unit refers to a single neuron and $ \varphi (\mathcal{H}) = 1 \, \text{if max}(\mathcal{H}) \geq t \, \text{else} \, \varphi (\mathcal{H}) = 0 $, where $t$ is a user-defined threshold, the function value of $ \varphi$ will be 1 as long as any of the output vectors of the neuron exceed the threshold $t$, indicating that the neuron is activated. Therefore, $\text{coverage}(f,\mathcal{D})$ represents the total number of activated neurons. This represents the most classic form of neuron coverage ~\cite{deepxplore}.
Alternatively, when a unit refers to a combination of all neurons within a network layer and $\varphi(\mathcal{H})=||\text{cov}(\mathcal{H})||$, where $\text{cov}(\mathcal{H})$ represents the covariance matrix, $\text{coverage}(f,\mathcal{D})$ represents the L1 norm of the covariance matrices across all network layers. This is known as NLC ~\cite{nlc}, the state-of-the-art neuron coverage criteria, which adopts $ \sum_{j=1}^m{( \sum_{i=1}^m | \varsigma_{n_{i}, n_{j}} | )} $ to represent the DNN
coverage, where m is the number of neurons and $ \varsigma_{n_{i}, n_{j}} = \mathbb{E}[(n_{i} - \mathbb{E}[n_i])(n_j - \mathbb{E}[n_j])] $ denotes the correlation between neuron $ n_i $ and $ n_j $.

However, it is worth noting that the majority of research efforts have primarily focused on image-based tasks. Only DeepCruiser~\cite{deepcruiser} and NLC have made some attempts to process text-related tasks, and these have predominantly been limited to simpler models such as RNNs and LSTMs. As a result, for transformer-based language models, which have gained popularity in recent years, the application of neuron coverage criteria remains largely unexplored.

\input{figure/figure2-2}

\subsection{Transformer-based Language Model}\label{section2-2}
In contrast to conventional RNN and CNN architectures, the transformer architecture, known for its self-attention mechanism ~\cite{vaswani2017attention}, effectively captures contextual relationships among words and tokens in text. Consequently, models designed based on this structure have achieved state-of-the-art performance in numerous language tasks and are commonly referred to as transformer-based language models. 
As shown in Figure \ref{figure2-2}, the self-attention mechanism is the core technology of the transformer architecture. The self-attention block computes the representation of each token by considering its relationship with all other tokens in the sequence. This is achieved using query (\(q\)), key (\(k\)), and value (\(v\)) vectors, where the attention score is calculated as:
\[ \text{Attention}(q, k, v) = \text{softmax}\left(\frac{qk^T}{\sqrt{d_k}}\right)v, \] 
with \(d_k\) being the dimension of the key vectors.
The transformer architecture includes two important modules: the encoder and the decoder, both centered around the self-attention block. 
The encoder, as an auto-encoding module, encodes the input into a latent representation, while the decoder, as an auto-regressive module, generates each token in the sequence one at a time, conditioning on the previously generated tokens. Practitioners can choose to use only the encoder, only the decoder, or both and add optional task-specific layers to form transformer-based language models. These choices represent three classic transformer-based model architectures.

In recent years, the transformer architecture has seen increased adoption due to its outstanding performance across various fields. 
Transformer-based language models are widely integrated into various software applications used in daily work and life, necessitating high safety standards and improved testing to further enhance model quality. 
These models have two key characteristics: large-scale model size and the self-attention mechanism, which make them more complex and harder to understand compared to traditional models. Many testing methods have been proposed for these models, but most are variants of traditional testing approaches, primarily black-box testing, and lack awareness of the internal neuron behavior within the models, i.e., they lack testing attempts guided by neuron coverage criteria.

%In particular, it is unclear whether neuron coverage testing is effective for these models. 
%Identifying suitable testing metrics and establishing effective testing methods is an urgent need.

%\jiayang{these sentences are about the motivation of this work, which is kind of duplicated with Section 2.3. Therefore, maybe consider adding more background info about transformer-based language morels (or LLMs) to mention their diverse task-handling capabilities and impacts on research and industry.}

\input{figure/figure2-3}

\subsection{Motivation and Our Study}
 \noindent\textbf{Motivation.}
The transformer architecture has revolutionized deep learning, particularly in natural language processing, leading to widespread societal and everyday applications. Ensuring the safety and security of these models is now a critical concern, presenting urgent challenges in testing and improvement.
Neuron coverage testing, primarily designed for image processing models like DNNs and CNNs, has shown its value in various applications. Therefore, it is essential to explore and comprehensively evaluate neuron coverage criteria for testing transformer-based language models.
Although there have been a few empirical studies on neuron coverage criteria, these studies have primarily evaluated older models and tasks, lacking recent state-of-the-art benchmarks. Furthermore, these studies have focused on traditional networks (excluding transformer architectures) and image tasks (with minimal focus on text tasks). Neuron coverage testing for transformer-based language models remains largely unexplored. This presents a pressing research question: To what extent are neuron coverage criteria effective for transformer-based language models, and how can their effectiveness be comprehensively evaluated?
%thus, it becomes challenging to consider the unique characteristics of transformers to propose effective coverage testing techniques that aid in enhancing the model.

\noindent\textbf{Our study.}    
To bridge the aforementioned research gap, we conduct an empirical study to evaluate the effectiveness of coverage criteria on transformer-based language models as shown in Figure ~\ref{figure2-3}. Since this study requires exploring internal neuron information, the target models must be open source. As noted in Section ~\ref{section2-2}, the core modules of transformers are the encoder and decoder. Therefore, we evaluate widely-used language models encompassing encoder-only (e.g., BERT, RoBERTa, Albert), decoder-only (e.g., GPT-2, LLaMa), and encoder-decoder (e.g., T5) architectures, covering all types of transformers.
Given the diverse coverage criteria available, we first perform a detailed analysis and summary of these criteria based on coverage unit granularity and algorithm complexity. Considering the large parameter sizes of transformer-based language models, only polynomial-time algorithms are feasible. As mentioned in Section ~\ref{section2-1}, neuron coverage includes various granularities. Therefore, we evaluate the coverage criteria, including the latest state-of-the-art criteria, covering granularities such as single neuron (indicated by circles), intra-layer neuron combination (indicated by diamonds), and inter-layer neuron combination (indicated by squares).
Ultimately, we conduct evaluations from three distinct aspects: Error-Revealing Capability and Diversity, which are related to test suite properties, and Input Generation Guidance, which pertains to test suite generation, unraveling the efficacy of neuron coverage criteria on transformer-based language models.

%% file: figure/figure2-1.tex
\begin{figure}[t]
	\centering
	\includegraphics[scale=0.65]{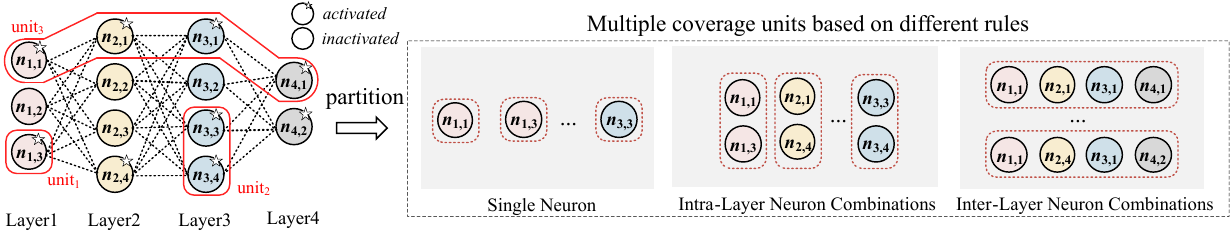}
   
	\caption{ Three forms of neuron coverage: single neuron, intra-layer neuron combination, and inter-layer neuron combination. The red circles highlight $\text{unit}_{1}$, $\text{unit}_{2}$ and $\text{unit}_{3}$, which are examples of  three forms, respectively.}
	\label{figure2-1}
\vspace{-5pt}
\end{figure}

%% file: figure/figure2-2.tex
\begin{figure}[t]
	\centering
	\includegraphics[scale=0.65]{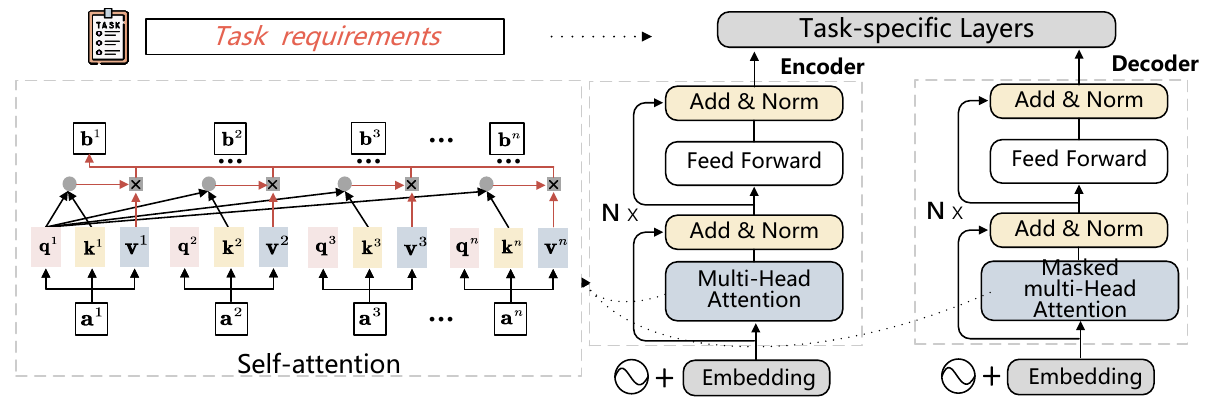}
	\caption{A transformer-based language model can be constructed using an encoder, a decoder, or a combination of both, along with task-specific layers. The encoder typically employs Multi-Head Attention, while the decoder utilizes Masked Multi-Head Attention.}
	\label{figure2-2}
 \vspace{-5pt}
\end{figure}

%% file: figure/figure2-3.tex
\begin{figure}[t]
	\centering
	\includegraphics[scale=0.12]{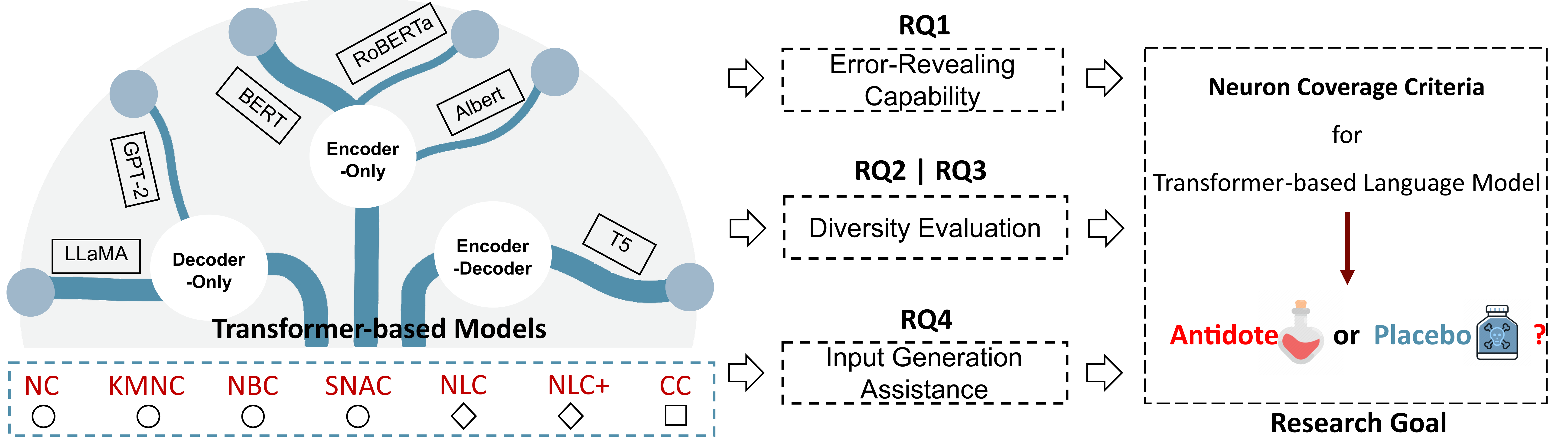}
    
	\caption{Comprehensive evaluation of neuron coverage criteria on transformer-based language models across three key aspects.}
 \label{figure2-3}
 
\end{figure}

%% file: chapter/criteria.tex
\section{Neuron Coverage Review: Granularity \& Complexity}
\label{sec:review}

Since DeepXplore~\cite{deepxplore} introduced the concept of neuron coverage in 2017, numerous neuron coverage criteria have been developed. These criteria encompass extensive structural coverage and limited non-structural coverage, with structural coverage being the predominant type. Structural coverage assesses the coverage of specific elements within a DNN during testing, primarily focusing on various forms of neuron coverage. In contrast, non-structural coverage ignores structural elements and instead evaluates the difference in DNN behaviors between a test input and the training data, termed surprise adequacy (SA). 
Neuron coverage and its extended versions are classified under structural coverage, which is the focus of this study. Despite numerous works introducing these criteria, a systematic summary is lacking. This chapter aims to provide a detailed analysis of various criteria, considering \emph{ coverage unit granularity} and \emph{algorithm complexity}. 
Next, we will use Figure~\ref{figure3-1} as an illustrative example to conduct a detailed analysis of these two aspects. Finally,  we will identify and select feasible and representative neuron coverage criteria for our evaluation.

\input{figure/figure3-1}
% Please add the following required packages to your document preamble:
% \usepackage{multirow}
%The first column specifies the category of each coverage metric, and the second column lists the coverage metrics. The third and fourth columns indicate the target models and datasets used in the original papers for these coverage metrics. The fifth and sixth columns respectively provide the storage complexity and computational complexity for each metric.

\vspace{5pt}
\noindent\textbf{Symbols and Notations.} 
Given a neural network \( N \) with \( l \) layers \(\{L_1, L_2, \ldots, L_l\}\), where each layer has \(\{n_1, \ldots, n_l\}\) neurons respectively, the total number of neurons is  \(m = \sum_{i=1}^{l} n_i= n_1 + \cdots + n_k \), and the total number of edges is \( e \leq \sum_{i=1}^{l-1}n_i n_{i+1} =  n_1 n_2 + \cdots + n_{l-1} n_l \). The training set is denoted as \(\mathcal{T}\), and the test suite is denoted as \(\mathcal{S}\), with \(|\mathcal{T}|\) and \(|\mathcal{S}|\) representing the sizes of the datasets. 
Given a sample, the computational cost for one inference by the neural network \( N \) is denoted as \( T_{\text{inference}} \).
Typically, the cost of neuron coverage criterion consists of two parts: 
1). Performing one inference for all samples in the test suite, with a computational cost of \(|\mathcal{S}| \times T_{\text{inference}}\) and a memory cost of \(|\mathcal{S}| \times m\).
2). The cost of computing the coverage criterion.
For the first part, the time cost is generally acceptable and is the same for each criterion. Therefore, our focus will be on the complexity analysis of the second part.

\input{table/table_criteria}

\begin{compactitem}[$\bullet$]
	
	\item \textbf{NC/NC+ (\textit{Neuron Coverage})}: 
	This criterion represents the proportion of activated neurons among all neurons. NC measures the proportion of neurons with activation greater than 0, while NC+ measures the proportion with activation above a certain threshold. Each neuron needs to record a boolean variable, i.e., whether it is activated, resulting in a memory complexity of \(\Theta(m)\). Calculating activation is a constant time operation, but this must be performed for each sample in the test suite. Consequently, the computational complexity is \(|\mathcal{S}| \times \Theta(m) = \Theta(|\mathcal{S}| \times m)\).
	
	\item \textbf{KMNC (\textit{K-Multisection Neuron Coverage})}: 
	First, the maximum and minimum values for each neuron are initialized using the training set, introducing an additional computational complexity of \(\Theta(|\mathcal{T}| \times m)\) and a storage complexity of \(\Theta(2 \times m)\). Each neuron is then divided into \(k\) intervals, and the proportion of activated intervals (i.e., the presence of at least one sample with an activation value within the interval) is calculated. Since checking if a value falls within an interval is a constant time operation, the memory complexity comes from the initialization \(\Theta(2 \times m)\) and the activation record for each interval \(\Theta(k \times m)\), resulting in \(\Theta((2+k) \times m)\). The computational complexity is \( \Theta(|\mathcal{S}| \times k \times m) + \Theta(|\mathcal{T}| \times m)\).
	
	\item \textbf{NBC (\textit{Neuron Boundary Coverage}) / SNAC (\textit{Strong Neuron Activation Coverage})}: 
	The initialization is the same as in KMNC. The conditions for the two criteria differ: NBC determines whether the activation value of a neuron is below the minimum or above the maximum, whereas SNAC only checks if it exceeds the maximum. Both operations are of constant time complexity. Therefore, the memory complexity includes the initialization \(\Theta(2 \times m)\) and the condition checks \(\Theta(m)\), resulting in a total complexity of \(\Theta(3 \times m)\). The computational complexity is \( \Theta(|\mathcal{S}| \times 2 \times m)+ \Theta(|\mathcal{T}| \times m)\) for NBC and \( \Theta(|\mathcal{S}| \times m)+ \Theta(|\mathcal{T}| \times m)\) for SNAC.
	
	\item \textbf{TKNC (\textit{Top-k Neuron Coverage})/ TKNP (\textit{Top-k Neuron Patterns})}: 
	Given the parameter \(k\), determine whether each neuron in a layer has an activation value among the top \(k\). This requires sorting the activation values of all neurons in a single network layer. Each neuron needs to record a boolean variable, so the memory complexity is \(\Theta(m)\). In addition, we also store the top-\(k\) activation patterns for each input. Each pattern includes \(k\) neurons, and aggregating across all layers leads to a storage complexity of \(\Theta\left(k \cdot \sum_{i=1}^{l} \binom{n_i}{k} \right)\), where \(n_i\) is the number of neurons in the \(i\)-th layer and \(l\) is the total number of layers.
	Since the sorting complexity for each layer is \(\Theta(n \log n)\) and there are \(l\) layers, the computational complexity is \(|\mathcal{S}| \times \Theta(n_1 \log n_1 + n_2 \log n_2 + \cdots + n_l \log n_l) = \Theta(|\mathcal{S}| \times \sum_{i=1}^{l}n_i log n_i)\).
	
	\item \textbf{NLC/NLC+ (\textit{NEURAL COVERAGE})}:
	NLC treats each neuron as a variable and computes the covariance matrix of multiple variables based on the outputs of all samples in each layer. Initially, it calculates the mean and then the covariance, with a computational complexity of \(\Theta(|\mathcal{S}| \cdot m) + O\left(|\mathcal{S}| \cdot \sum_{i=1}^{l}\frac{n_i(n_i+1)}{2}\right)\). As a result, each layer must store a covariance matrix, leading to a memory complexity of \(O(n_1^2 + n_2^2 + \ldots + n_l^2) = O\left(\sum_{i=1}^{l}n_i^2\right)\). The total computational complexity is \(\Theta(|\mathcal{S}| \cdot m) + O\left(|\mathcal{S}| \cdot \sum_{i=1}^{l}\frac{n_i(n_i+1)}{2}\right)\). NLC+ uses an incremental calculation method, adding samples only when the Frobenius norm of the covariance matrix increases, thus achieving a slightly lower overall complexity compared to NLC, but of the same order of magnitude.
	
	\item \textbf{CC$_w$ (\textit{Causality-aware Coverage})}: Firstly, the neural network is abstracted as a causal graph \(G=(V,E)\) where \(|V|=m\) and \(|E|=e\), which can be directly derived from the DNN structure. Directed edges may exist between any two neurons in adjacent layers, leading to a memory complexity of \(\Theta(m + 2e)\). Then, the causal graph \(G=(\hat{V}, \hat{E})\) is extracted and recovered by the test suite. The CC metric calculates the proportion of edges recovered, i.e., \(|E \cap \hat{E}|/|E|\).
	For each edge \((v_1,v_2)\), it is determined if no subset \(z\) belonging to \(V\) exists such that \(v_1\) and \(v_2\) are independent conditioned on \(z\). If so, the edge exists. The specific procedure is as follows: assume the target edge is \(x \rightarrow y\). Considering \(\{v_1, v_2, \ldots, v_k\}\) as the confounding neurons for this edge, the structural assignment to \(y\) is given by \(f_y(x, v_1, \ldots, v_k) \rightarrow y\).
	Then, for each neuron \(v_i \in \{v_1, \ldots, v_k\}\), compute \(\alpha_i \leftarrow \frac{|\text{cov}(v_i, x)|}{\sum_{j=1}^{k} |\text{cov}(v_j, x)|}\) and \(v'_i \leftarrow (1 - \alpha_i)v_i\). This results in \(y' \leftarrow f_y(x, v'_1, \ldots, v'_k)\), and an independence test is conducted between \(x\) and \(y'\).
	The covariance matrix needs to be stored, with a memory complexity similar to NLC, \(O\left(\sum_{i=1}^{l}n_i^2\right)\). Additionally, for each edge, updating \(y'\) requires \(2 \times O(n_i)\), where \(n_i\) is the number of neurons in a layer, approximated by \(m/l\). The independence test complexity is approximately \(O(K \times m)\), where \(K\) is a statistical parameter. 
	 Note that it is sufficient to select only a subset of network layers for analysis. Let $w$ denote the number of selected layers. The total computational complexity is roughly \(\Theta(|\mathcal{S}| \cdot m) + O\left(|\mathcal{S}| \cdot \sum_{i=1}^{w}\frac{n_i(n_i+1)}{2}\right) + \Theta(e) + \Theta(K \times m)\).
	Prior work has shown that selecting only three layers is sufficient for achieving effective coverage~\cite{ji2023cc}, and we accordingly set $w=3$ in this study. 
	
	\item \textbf{CTC (\textit{Combinatorial Testing Coverage})}:
	Considering \(t\)-way combinatorial dense coverage, each neuron has two possible states \(\{0, 1\}^{2}\), resulting in \(\binom{n_{i}}{t}\) combinations in the \(i\)-th layer. For each \(t\)-way neuron combination, the states are calculated, leading to computational complexity of \(|\mathcal{S}| \times t \times O\left(\binom{n_{1}}{t} + \binom{n_{2}}{t} + \cdots + \binom{n_{l}}{t}\right)\). Considering that each neuron combination has \(2^{t}\) possible states, the total number of states is \(2^{t} \cdot \Theta\left(\binom{n_{1}}{t} + \cdots + \binom{n_{l}}{t}\right)\). Thus, the computational complexity is \(|\mathcal{S}| \times t \times \Theta\left(\binom{n_{1}}{t} + \binom{n_{2}}{t} + \cdots + \binom{n_{l}}{t}\right)\). The memory complexity is \(|\mathcal{S}| \times 2^t \times \Theta\left(\binom{n_{1}}{t} + \binom{n_{2}}{t} + \cdots + \binom{n_{l}}{t}\right)\).

	\item \textbf{SS (\textit{Sign-Sign Coverage})}: 
	Using Sign-Sign Cover as the coverage method, we consider two neurons \(a\) and \(b\) in adjacent layers \(L_i\) and \(L_{i+1}\). The coverage checks if \(a\) independently influences the sign change of \(b\); i.e., there exist two samples \(x_1\) and \(x_2\) such that all neurons in layer \(L_i\) except \(a\) remain unchanged, while the sign of neuron \(b\) in layer \(L_{i+1}\) changes.
	The memory complexity is determined by the total number of edges \(e\), as we need to store information about whether each pair of adjacent neurons satisfies the coverage criterion. Determining coverage involves selecting two samples from the test suite, with a complexity of \(\Theta(|\mathcal{S}| \times (|\mathcal{S}|-1))\). Then, we check if the sign of each neuron changes \(\Theta(m)\), and verify if only one neuron per layer undergoes a sign change \(\Theta(m)\). Thus, the memory complexity is \(\Theta(m + e)\), and the total computational complexity is \(\Theta(|\mathcal{S}| \times (|\mathcal{S}|-1) \times m^2)\).

	\item \textbf{Pcov (\textit{Path-driven Coverage})}: 
	Using Full State Path Coverage as an example, a \(k\)-path is defined as a path of length \(k\) formed by neurons from \(k\) consecutive layers. There are a total of \(\sum_{i=1}^{l-k+1} \left( \prod_{j=i}^{i+k-1} n_j \right)\) paths. For each path, each neuron can either be activated (1) or not (0), resulting in \(2^k\) possible states for each path. Therefore, the memory complexity is \(\Theta(\sum_{i=1}^{l-k+1} \left( \prod_{j=i}^{i+k-1} n_j \right) \times 2^k)\).
	For each sample, the state of each path is calculated and activated, with a computational complexity of \(|\mathcal{S}| \times \Theta (\sum_{i=1}^{l-k+1} \left( \prod_{j=i}^{i+k-1} n_j \right) \times k) \).

	\item \textbf{NPC (\textit{Neuron Path Coverage}) / IDC (\textit{Importance-Driven Coverage})}: 
	Building on prior research, NPC / IDC necessitates a white-box approach to DNNs, specifically requiring layer-wise relevance propagation techniques. This requirement is often impractical in testing scenarios. Additionally, its core mechanism of layer-wise propagation is constrained to discriminative models. Consequently, it is not included in this analysis and comparison.
\end{compactitem}

\noindent \textbf{Summary.} 
Based on the comprehensive analysis presented and as detailed in the Table ~\ref{criteria}, coverage unit granularity can be categorized into three levels: single neuron (NC/NC+, KMNC, NBC, SNAC, IDC), intra-layer neuron combination (TKNC, TKNP, NLC/NLC+, CTC), and inter-layer neuron combination (CC$_w$, SS, Pcov, NPC). Regarding algorithmic complexity, these methods can be distinguished between those exhibiting exponential complexity (Pcov, CTC) and those with polynomial complexity (others). Additionally, considering the requirement for gradient calculation, they can be classified as either white-box (NPC, IDC) or non-white-box (others).
The analysis indicates that while TKNC, TKNP, SS, NPC, and IDC have certain limitations, TKNC, TKNP, and SS, despite not exhibiting exponential complexity, still demonstrate higher complexity compared to most other coverage criteria.
Consequently, we have selected NC, KMNC, NBC, SNAC, NLC, NLC+, and CC$_w$. These criteria encompass all coverage unit granularities, forming a representative set that includes two state-of-the-art (SOTA) coverage criteria which have not been evaluated in other studies.

%% file: figure/figure3-1.tex
\begin{figure}[t]
	\centering
	\includegraphics[scale=0.3]{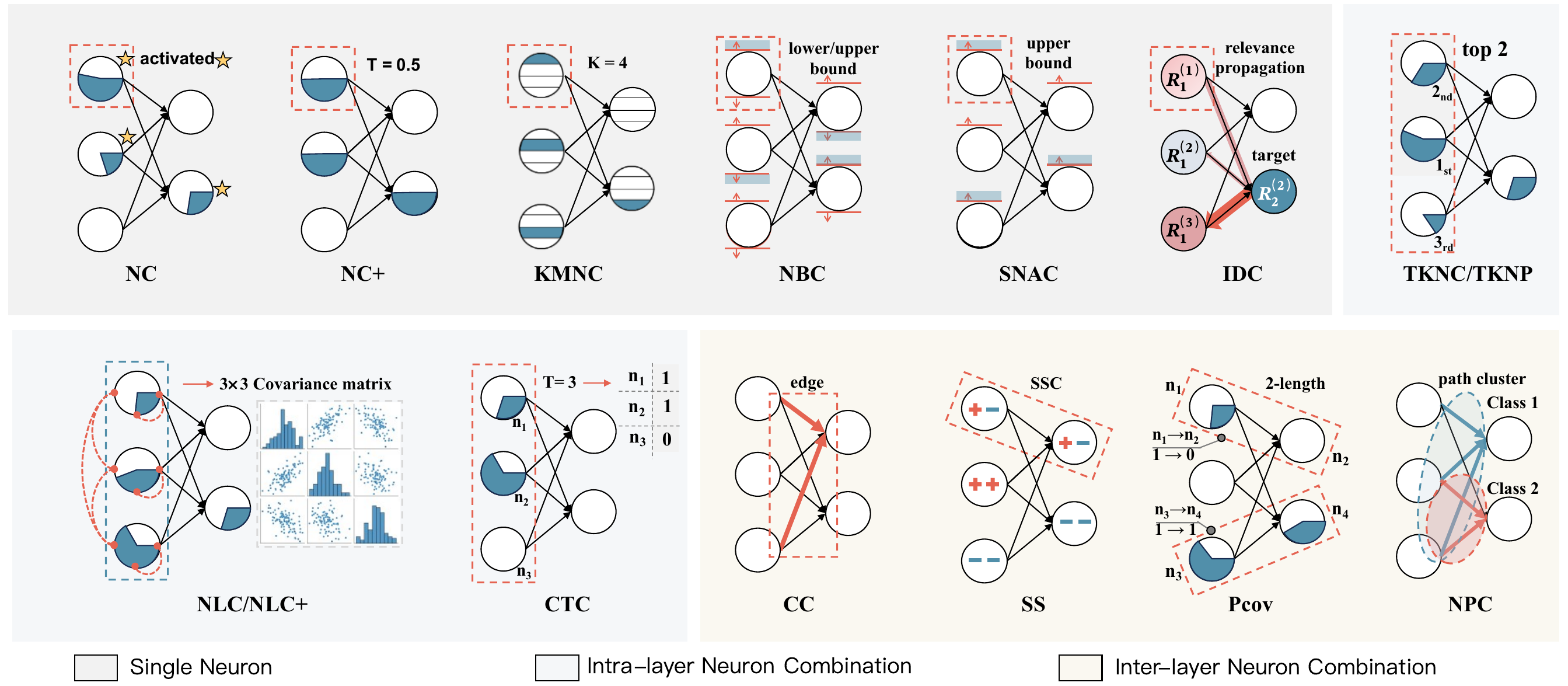}
	\caption{An illustration depicting all neuron coverage criteria. From top to bottom, left to right, it sequentially presents the intuitive visual representation of three parts (with distinct background colors): single neuron, intra-layer neuron combination, and inter-layer neuron combination criteria.}
	\label{figure3-1}
\vspace{-5pt}
\end{figure}

%% file: table/table_criteria.tex
\begin{table}[]
	\caption{Here are the details of each coverage criterion, including the subject models, datasets, memory complexity, and computational complexity. The "\CheckmarkBold" and "\XSolidBrush" symbols in the last two columns indicate whether the complexity is polynomial and whether the criteria are used in this paper. The |$\mathcal{T}$| and |$\mathcal{S}$| represent the sizes of the training set and test suite, respectively. The $m$, $l$, and $e$ represent the number of neurons, layers, and edges in the target model, respectively, and $n_{i}$ represents the number of neurons in the $i$-th layer.}
	\label{criteria}
	\renewcommand{\arraystretch}{1.5}
	\centering
	\resizebox{\textwidth}{!}{
		\begin{tabular}{llllllll}
			\hline
			\textbf{Coverage Unit}                                                                             & \textbf{Coverage Criteria}                                  & \textbf{Model }                 & \textbf{Dataset}           & \textbf{Memory Complexity} & \textbf{Computation Complexity} & \textbf{Polynomial} & \textbf{Used} \\ \hline
			\multirow{5}{*}{\textbf{Single Neuron}}                                                            & NC~\cite{deepxplore} /NC+~\cite{deepgauge}     & \makecell[l]{\rule[-5pt]{0pt}{17pt}FCN CNN\\ ResNet} & \makecell[l]{\rule[-5pt]{0pt}{17pt}Image PDF\\ Android} & $\Theta(m)$     & $\Theta(|\mathcal{S}| \times m)$  & \CheckmarkBold    & \CheckmarkBold    \\
			& KMNC~\cite{deepgauge}        & CNN ResNet             & Image             & $\Theta((2+k) \times m$) & $\Theta(|\mathcal{S}|\times k \times m)+\Theta(|\mathcal{T}|) \times m)$    & \CheckmarkBold          & \CheckmarkBold    \\
			& NBC~\cite{deepgauge}         & CNN ResNet             & Image             & $\Theta(3 \times m)$                 & $\Theta(|\mathcal{S}| \times 2 \times m) + \Theta(|\mathcal{T}| \times m)$     & \CheckmarkBold          & \CheckmarkBold    \\
			& SNAC~\cite{deepgauge}        & CNN ResNet             & Image             & $\Theta(3 \times m)$                 & $\Theta(|\mathcal{S}| \times m) + \Theta(|\mathcal{T}| \times m)$                      &  \CheckmarkBold          & \CheckmarkBold   \\
			& IDC~\cite{importance} & CNN  & Image & ------  & ------    & ------          & \XSolidBrush    \\ \hline
			\multirow{4}{*}{\makecell[l]{\rule[-5pt]{0pt}{25pt}\textbf{Intra-layer} \\ \textbf{Neuron}\\ \textbf{Combination}}} 
			& TKNC/TKNP~\cite{deepgauge}       & CNN ResNet             & Image             &  $\Theta(k \cdot \sum_{i=1}^{l} \binom{n_i}{k})$  &  $\Theta(|\mathcal{S}| \times \sum_{i=1}^{l} n_{i}logn_{i})$                   & \CheckmarkBold          & \XSolidBrush    \\
			& NLC/NLC+~\cite{nlc}          & \makecell[l]{\rule[-5pt]{0pt}{17pt}CNN MobileNet\\ LSTM GAN} & Image Text        & $\Theta(\sum_{i=1}^{l} n_{i}^{2})$  & \makecell[l]{\rule[-5pt]{0pt}{17pt}$O(|\mathcal{S}| \times \sum_{i=1}^{l} \frac{n_{i}(n_{i}+1)}{2}) +$ \\ $ \Theta(|\mathcal{S}|) \times m + O(\sum_{i=1}^{l} n_{i}^{2})$}  & \CheckmarkBold          & \CheckmarkBold    \\
			& CTC~\cite{deepct}            & FCN                    & Image             & $|\mathcal{S}| \times 2^{t} \times \Theta(\binom{n_{1}}{t} + ... + \binom{n_{l}}{t})$    & $|\mathcal{S}| \times t \times \Theta(\binom{n_{1}}{t} + ... + \binom{n_{l}}{t})$   & \XSolidBrush          &  \XSolidBrush    \\ \hline
			\multirow{4}{*}{\makecell[l]{\rule[-5pt]{0pt}{17pt}\textbf{Inter-layer} \\ \textbf{Neuron}\\ \textbf{Combination}}} & CC$_w$~\cite{ji2023cc} & \makecell[l]{\rule[-5pt]{0pt}{17pt}ResNet VGG\\ LeNet}       & Image             &  $\Theta(\sum_{i=1}^{w} n_{i}^{2})$  &  \makecell[l]{\rule[-5pt]{0pt}{17pt}$\Theta(e) + O(|\mathcal{S}| \times \sum_{i=1}^{w} \frac{n_{i}(n_{i}+1)}{2}) + $\\ $\Theta(|\mathcal{S}| \times m) + \Theta(K \times m)$ }  & \CheckmarkBold     & \CheckmarkBold    \\
			& SS~\cite{deepcover}          & FCN   & Image & $\Theta(m+e)$  & $\Theta(|\mathcal{S}| \times (|\mathcal{S}|-1) \times m^{2})$   & \CheckmarkBold  & \XSolidBrush    \\
			& Pcov~\cite{deeppath}         & FCN                    & Image             & $\Theta(\sum_{i=1}^{l-k+1} \Big( \prod_{j=1}^{i+k-1} n_{j} \Big) \times 2^{k})$    &  $|\mathcal{S}| \times \Theta(\sum_{i=1}^{l-k+1} \Big( \prod_{j=1}^{i+k-1} \Big) \times k)$  & \XSolidBrush         &  \XSolidBrush   \\
			& NPC~\cite{npc}               & CNN                    & Image             & ------         & ------              & ------     & \XSolidBrush    \\ \hline
	\end{tabular}}
\vspace{-10pt}
\end{table}

%% file: chapter/empirical.tex
\vspace{-6pt}
\section{Study Design And findings}
\label{sec:study_design}

\subsection{Research Questions}

In response to the identified research gap, we conduct a comprehensive study to assess the effectiveness of neuron coverage criteria in testing transformer-based language models.
We approach the problem from the following three perspectives: \textit{error-revealing capability} (RQ1), \textit{diversity} (RQ2, RQ3), and \textit{input generation guidance} (RQ4).
First, we investigate whether these criteria could help the model identify error samples. By introducing various types of noise to a sample, we classify the resulting samples as either adversarial, which induces classification errors, or benign, which does not affect classification. Adversarial samples significantly contribute to the coverage criteria of the test suite compared to benign samples(RQ1).
Second, we examine whether these criteria could assess the diversity of the test suite, focusing on distribution diversity and sample fairness. The first aspect analyzes whether a more dispersed distribution leads to greater coverage(RQ2), while the second aspect evaluates whether more balanced categories result in higher coverage(RQ3).
Finally, we explore whether these criteria could facilitate sample generation. By incorporating coverage guidance into the standard sample generation process, we aim to determine if the generated samples could uncover a higher proportion of error corner cases, such as adversarial samples(RQ4).
The specific research questions are as follows:

\begin{compactitem}[$\bullet$]
	\item \textbf{RQ1 (Error-Revealing Capability):} To what extent do neuron coverage criteria reveal the error-revealing capability of a test suite?
	
	\item \textbf{RQ2 (Distribution Diversity):} Can neuron coverage criteria evaluate the distribution diversity of a test suite?
	
	\item \textbf{RQ3 (Sample Fairness):} Can neuron coverage criteria assess the sample fairness within a test suite?
	
	\item \textbf{RQ4  (Input Generation Assistance):} Can coverage criteria guide fuzz testing to produce higher-quality test suites?
\end{compactitem}

%Finally, ERQ3 investigates test diversity, employing the number of covered classes as a measure for one aspect of diversity and the distribution of %diversity for another. This comprehensive approach aims to provide a nuanced understanding of the performance and applicability of coverage criteria on %transformer-based language models. 
%The transformer architecture, primarily based on self-attention mechanisms, differs substantially from their CNN and RNN counterparts.
%Thus,
%\textcolor{blue}{We conduct an empirical study with the following empirical research questions(ERQs) to investigate the applicability of traditional criteria to transformer-based neural networks.}
\subsection{Experiment Setup} 
\input{table/table_TLM}
\input{table/table_Dataset}
\subsubsection{Target Models and Datasets}
From Section ~\ref{sec:review}, we understand that current research predominantly focuses on DNN and CNN models, mainly in image tasks. Hence, we will focus on transformer models for text tasks, addressing gaps in datasets and models from previous studies.
Since coverage analysis requires accessible model information and manageable computational costs, our investigation targets three distinct architectures of transformer-based language models: encoder-only models (such as BERT~\cite{devlin2018bert}, RoBERTa, and ALBERT~\cite{lan2019albert}), encoder-decoder models (such as T5), and decoder-only models (like GPT-2 and LLaMA). All these models are derived from the transformer architecture (see Table ~\ref{tab:model_summary}).
These models are open-source and widely used, with parameter scales ranging from small to large, covering all types of transformer models. 
To comprehensively assess their capabilities, we select these models as our targets and conduct experiments on six tasks in total. Among them, four are classification tasks, namely Sentiment Analysis, Semantic Matching, Natural Language Inference, and Text Classification, which correspond to the AGNews, IMDB, QQP, and SNLI datasets. In addition, we consider two generation tasks, Machine Translation and Summarization, which are evaluated on the WMT14 and Gigaword datasets, respectively (see Table~\ref{tab:data_summary}).
To ensure robustness in our findings, each configuration underwent five training iterations. For context, we present text length statistics for these datasets in Table ~\ref{tab:data_summary}. Notably, the IMDB dataset has an average text length of 233 words, significantly longer than the other datasets, where the average length is less than 40 words. Consequently, we set the maximum text lengths to 256 for IMDB and 128 for the other datasets. These settings allow the target models to achieve comparable performance and effectively assimilate information from the inputs.

%As shown in figure, all configurations include architecture(BERT, ALBERT, ROBERTA), Dataset(IMDB, AGNEWS) and criteria(NC, KMNC, NLC). 
\subsubsection{Target Neuron Coverage Criteria}
Considering that our analysis focuses on large-scale, transformer-based language models, we choose algorithms with polynomial time complexity to ensure computations are completed within an acceptable timeframe. To ensure representativeness, we select a set of neuron coverage criteria across three granularities: NC/NC+, KMNC, NBC, SNAC, NLC, NLC+, and CC.

\subsubsection{Implementations}
\label{settings}
Our experimental framework consists of three primary implementation components:
\noindent \textbf{1): Neuron Coverage Criteria}:  While the criteria we employ are supported by open-source code, significant modifications were necessary to standardize them within the PyTorch framework and to ensure their compatibility with transformer architectures. We meticulously summarize and refactor the existing criteria, enhancing their applicability and accuracy for our specific use case. The implementation involves two critical aspects: the definition of neurons and the calculation of coverage.

\begin{compactitem}[$\bullet$]
	\item 
\uline{Neuron definition in Transformers.}
	Transformer models consist of multiple encoder or decoder blocks, each containing a multi-head attention (MHA) module and a feed-forward network (FFN). 
	In our implementation, neurons are defined as the outputs of both the MHA and FFN modules. 
	Specifically, for an input $X \in \mathbb{R}^{n \times d}$, the MHA produces
	\[
	\text{MHA}(X) = \text{Concat}(h_1, \ldots, h_H)W^O,
	\]
	where $\text{Concat}(h_1, \ldots, h_H)$ denotes the concatenated head outputs, and $W^O$ is the output projection matrix. 
	We treat both the concatenated head representations and their projected outputs as neuron activations. 
	For the FFN, consisting of two fully connected layers,
	\[
	\text{FFN}(x) = \sigma(xW_1 + b_1)W_2 + b_2,
	\]
	we consider both the intermediate activation $\sigma(xW_1 + b_1)$ and the final output as neurons. 
	In contrast, intermediate variables such as the query, key, and value ($Q,K,V$) are excluded, since they serve as latent factors for attention computation rather than independent activation units. 
	This definition ensures consistency with the representational semantics of neural network layers in existing coverage criteria.

	\item 
	\uline{Coverage calculation in Transformers.}
	Transformer models process sequences of tokens in a manner analogous to the shared-kernel mechanism in CNNs: while CNNs slide convolutional kernels over spatial dimensions, transformers apply the same parameters across tokens in the sequence. Inspired by prior work on coverage testing for CNNs, where activations are commonly pooled across spatial locations within each channel, and by established testing practices in natural language processing~\cite{du2019deepstellar,zheng2019testing} that consider the entire input sequence, we adopt an averaging strategy over the token dimension and incorporate activations from all input tokens. This design not only reduces noise and enables efficient coverage computation by producing a stable sequence-level representation, but also ensures broad applicability across different Transformer architectures.

\end{compactitem}

\noindent 2): Transformer Model Fine-tuning: We leverage pre-trained models sourced from HuggingFace, upon which we integrate task-specific layers tailored to our experimental needs. Each model undergoes fine-tuning across four distinct tasks, meticulously calibrated to achieve comparable performance levels, thereby ensuring the reliability and validity of our results.

\noindent 3): Adversarial Attacks: To evaluate the robustness of our models, we employ the TextAttack API to conduct adversarial attacks. This involves the application of three diverse and sophisticated methods, allowing us to generate a representative set of adversarial samples and further test the sensitivity of our metrics to these adversarial examples.

\vspace{1ex}
\noindent \textbf{Implementation Environment}: All experiments were conducted on a Linux server running Ubuntu 20.04.4 LTS, equipped with an Intel(R) Xeon(R) Silver 4216 CPU, 376GB of RAM, and NVIDIA RTX A6000 GPUs. The implementation environment was built on Python v3.7.11. We implemented the neuron coverage criteria and model fine-tuning based on PyTorch v1.9.0 and the HuggingFace Transformers library v4.30.2. For the adversarial attack component, we utilized the TextAttack framework v0.3.9.

%% file: table/table_TLM.tex
\begin{table}[t]
\centering
\caption{Summary of Transformer-Based Language Models. This table provides an overview of notable transformer-based language models, including their developers, year of release, descriptions, architectures, parameter counts, and the number of layers.}
\label{tab:model_summary}
\resizebox{.95\textwidth}{!}{%
\begin{tabular}{@{}lllllll@{}}
\toprule
\textbf{Model} & \textbf{Developer} & \textbf{Year} & \textbf{Description} & \textbf{Architecture} & \textbf{Parameters} & \textbf{$\#$Layers}\\ \midrule
\textbf{BERT} & Google & 2018 & Pioneering transformer model in NLP, employing pre-training and fine-tuning. & Encoder-only & 110M & 12 \\
\textbf{ALBERT} & Google & 2019 & A lite version of BERT focusing on parameter-reduction techniques. & Encoder-only & 12M & 12  \\
\textbf{RoBERTa} & Facebook AI & 2019 & An optimized method for pre-training BERT models, with improved performance. & Encoder-only & 125M  & 12  \\
\textbf{GPT-2} & OpenAI & 2018 & A large language model and the second in the series of GPT models. & Decoder-only & 1.5B & 48 \\
\textbf{LLaMA} & Meta AI & 2023 & A collection of large language models, ranging from 7B to 65B parameters. & Decoder-only & 7B  & 32  \\
\textbf{T5} & Google & 2019 & Converts all NLP problems into a text-to-text format, trained using teacher forcing. & Encoder-decoder & 220M  & 12 \\
\bottomrule
\end{tabular}%
}
\vspace{-10pt}
\end{table}

%% file: table/table_Dataset.tex
\begin{table*}[t]
\normalsize
% \vspace{0.1cm}
\caption{Summary of Language Task Datasets. This table provides an overview of various language task datasets, detailing their training and test set sizes, maximum and average input lengths, tokenizer maximum lengths, number of classes, and the specific tasks they are used for.}
\label{tab:data_summary}
\centering

\resizebox{.95\textwidth}{!}{
\begin{tabular}{llllllll} \toprule
\textbf{Datasets} & \textbf{\#Train} & \textbf{\#Test} & \textbf{Input\_maxlen} & \textbf{Input\_avglen} & \textbf{Tokenizer\_maxlen} & \textbf{\#Classes} & \textbf{Task}\\ \hline
\textbf{AGNews} & 120,000 & 7,600 & 2470 & 233 & 256 & 4 & News Classification \\
\textbf{IMDB} & 25,000 & 25,000 & 203 & 39.8 & 128 & 2 & Sentiment Analysis \\
\textbf{QQP} & 364,000 & 40,000 & 270 & 22.2 & 128 & 2 & Question Pair \\
\textbf{SNLI} & 550,000 & 10,000 & 112 & 20.3 & 128 & 3 & Natural Language Inference \\ 

\textbf{gigaword} & 26,696& 2,967 & 2,937 & 21.3 & 128 & - & Machine Translation \\ 
\textbf{wmt14} & 4,508,785 & 3,000 & 112 & 20.3 & 128 & - & Summarization \\ 
\bottomrule
\end{tabular}}
\vspace{-10pt}
\end{table*}

%% file: chapter/evaluation1.tex
\subsection{Error-Revealing Capability ( RQ1 ) }

The error-revealing capability pertains to the ability of a test suite to uncover errors in a model when erroneous samples are incorporated. This research question seeks to explore the extent to which neuron coverage criteria can reflect the error-revealing capability of a test suite.  Specifically, it examines whether neuron coverage criteria can exhibit significant changes when erroneous samples are integrated into the test suite, indicating their sensitivity to newly introduced errors.

\subsubsection{Experiment Design} \vspace{0.5em}\noindent

To address RQ1, our experiment is divided into two phases:
1) Coverage Sensitivity to Error Samples: Evaluate whether the coverage criteria respond to erroneous samples. Specifically, does incorporating adversarial examples enhance coverage?
2) Factors of Coverage Increase: Identify whether the main reason for coverage increase is the addition of new samples or erroneous samples. 

To disentangle these two factors, we note that adversarial (erroneous) samples can be regarded as a special case of perturbed inputs where perturbations cause misclassification. Hence, to separate the error-related effect from the perturbation itself, we introduce noise samples (perturbed but non-adversarial) alongside the original and adversarial sets. Consequently, we construct three sets as follows:

\noindent \textit{Original Samples (denoted as $\mathcal{S}$):}
\[
\mathcal{S} = \left\{ x_i \in \text{Dataset}_{\text{test}} \mid f(x_i) = y_i \right\}; \quad \left| \mathcal{S} \right| = n; \quad \forall l \in \{1, 2, \ldots, c\}, \sum_{x_i \in \mathcal{S}} \mathbb{1}(y_i = l) = \frac{n}{c}.
\]
Here, $f(\cdot)$ denotes the neural network, $y_i$ the ground-truth label, and $c$ the number of classes. Thus, $\mathcal{S}$ is a class-balanced subset of correctly classified test samples, with $n=5000$ in our setup.
%This formula represents a sample set, which preserves a class-balanced subset of the test dataset. 
 %Here, $\text{Dataset}_{\text{test}}$ represents the test set, $x_i$ denotes a sample from the test set, and the total number of samples is $|\mathcal{S}| = n$. The neural network is denoted by $f(\cdot)$, and the number of classes in the dataset is $c$. The term $f(x_i)$ represents the predicted label, $y_i$ is the ground truth label of the sample, and $\mathbb{1}(y_i = j)$ equals 1 if $y_i = j$, otherwise it equals 0.  Therefore, these samples are correctly classified as benign samples with balanced class labels, meaning each class has an equal number of samples.
%In this experimental setup, the value of $n$ is set to 5000.

\noindent \textit{Noise Samples (denoted as $\mathcal{S}_{\text{noise}}$):}
\[
\mathcal{S}_{\text{noise}} = \left\{ \tilde{x}_i = \text{Perturb}(x_i) \mid x_i \in \mathcal{S}, \; f(\tilde{x}_i) = f(x_i) \right\}; \quad \left| \mathcal{S}_{\text{noise}} \right| = \left| \mathcal{S} \right|.
\]
Each $\tilde{x}_i$ is obtained by perturbing $x_i$ at either the character or word level, while ensuring the model prediction remains unchanged ($f(\tilde{x}i)=f(x_i)$). Hence, $\mathcal{S}{\text{noise}}$ has the same size and label distribution as $\mathcal{S}$.
%This formula represents a noise-perturbed sample set of the same size as the original samples. Here, $\text{Perturb}(\cdot)$ denotes the method for adding noise at the text level, and $\text{Perturb}(x_i)$ represents the noise sample obtained by perturbing $x_i$. The terms $f(\tilde{x}_i)$ and $f(x_i)$ represent the output values of the noise and original samples, respectively. For each $x_i \in \mathcal{S}$, the perturbation operation $\text{Perturb}(x_i)$ is performed, ensuring that $f(\tilde{x}_i) = f(x_i)$. Thus, the noise samples maintain the same classification as the original samples, resulting in a set $\mathcal{S}_{\text{noise}}$ of the same size, i.e., $\left| \mathcal{S}_{\text{noise}} \right| = \left| \mathcal{S} \right|$. In this experimental setup, the perturbation methods used are character-level and word-level. Character-level perturbation involves randomly adding, deleting, repeating, or replacing characters, while word-level perturbation involves similar operations on words.

\noindent \textit{Error Samples (denoted as $\mathcal{S}_{\text{adv}}$):}
\[
\mathcal{S}_{\text{adv}} = \left\{ \hat{x}_i = \text{AdvAttack}(x_i) \mid x_i \in \mathcal{S}, \; f(\hat{x}_i) \neq f(x_i) \right\}; \quad \left| \mathcal{S}_{\text{adv}} \right| = \left| \mathcal{S} \right|.
\]
Here, $\hat{x}_i$ is generated by applying text-level adversarial attacks to $x_i$, which intentionally alter the model prediction. We adopt character-level (DeepWordBug), word-level (TextFooler), and hybrid (TextBugger) attacks, consistent with the perturbation granularity used for $\mathcal{S}_{noise}$. For generation tasks, we further employ Seq2Sick and Morpheus as adversarial attack methods.

%This formula represents an adversarial sample set of the same size as the original samples. Here, $\text{AdvAttack}(\cdot)$ denotes the adversarial attack method at the text level, and $\text{AdvAttack}(x_i)$ represents the adversarial sample obtained by attacking $x_i$. For each $x_i \in \mathcal{S}$, the adversarial attack method $\text{AdvAttack}(x_i)$ is applied, ensuring that $f(\hat{x}_i) \neq f(x_i)$. Thus, the adversarial samples are derived from the original samples with altered classifications, resulting in a set $\mathcal{S}{\text{adv}}$ of the same size, i.e., $\left| \mathcal{S}_{\text{adv}} \right| = \left| \mathcal{S} \right|$. In this experimental setup, the adversarial attack methods employ include character-level (DeepWordBug \cite{deepwordbug}), word-level (TextFooler \cite{textfooler}), and a combination of character-level and word-level (TextBugger \cite{textbugger}) perturbations, consistent with the granularity of text-level perturbations used for noise samples.
\input{figure/figure8.tex}

Figure~\ref{fig:RQ1} provides specific examples using a sample from the QQP dataset. In this example, a square represents an original sample \( x_i \) belonging to the "not\_equivalent" class. The example distribution on the right shows that this sample falls correctly within the "not\_equivalent" region. Next, Perturbation 1, as an instance of \(\text{Perturb}(x_i)\), is a character-level alteration changing "How" to "Htow," transforming it into a circle representing a noise sample \(\tilde{x}_i\). This minor perturbation retains the sample in the "not\_equivalent" class, as shown in the right distribution example. However, Perturbation 2, as an instance of \(\text{AdvAttack}(x_i)\), changes "medicine" to "medicwne," resulting in a triangle representing an adversarial sample \(\hat{x}_i\). This perturbation achieves an adversarial attack by shifting the sample close to its original position but into the "equivalent" class range, as illustrated in the right distribution example.

Having established these three test suites, we design a two-phase experiment to evaluate their fault-revealing capabilities. 
The first phase, coverage sensitivity, assesses the benefit of increasing coverage by involving adversarial examples, a method utilized in several studies \cite{deepgauge,deepct,deeppath,deepcover}. Following previous work, we first introduce in-distribution samples (from either the train set or the test set) to the neural network and calculate the value of the given coverage criterion $\mathcal{C}$, denoted as $value\_\mathcal{S}$, representing the proportion of the neural network covered by the original samples. Next, we incorporate adversarial samples into the original samples and reintroduce this combined sample set to the neural network, calculating the coverage criterion $\mathcal{C}$ again, denoted as $value\_\mathcal{S}_{+adv}$. We then compare $value\_\mathcal{S}$ and $value\_\mathcal{S}_{+adv}$ to determine the difference in coverage between the original sample set and the augmented set with adversarial samples. This comparison reveals whether these slightly perturbed adversarial samples can increase coverage and expand the scope of the test suite. 
This step allows us to examine the impact of adversarial examples on coverage sensitivity and their potential to enhance coverage.
While this step allows us to confirm the sensitivity of the coverage criterion to adversarial samples, it overlooks the impact of test suite size. As the number of samples increases with the addition of adversarial examples, it remains unclear whether the rise in coverage is due to the erroneous information in the adversarial samples or simply the inclusion of additional samples.
A few studies have raised this concern \cite{ji2023cc,nlc,yan2024revisiting} and designed experiments to address it. 
Inspired by previous work, we have designed an ablation experiment to address this issue, specifically exploring the main factors contributing to coverage increase. Specifically, since error samples are generated by applying slight text-level perturbations to the original samples until misclassification occurs, we aim to investigate whether the main factor of coverage increase is the addition of new samples or the addition of adversarial samples. To this end, we created a noise sample set using perturbations similar to those employed in adversarial attacks. The only difference between this noise sample set and the error samples is whether the samples are correctly classified.
Thus, similar to the first phase, we introduce the combined set of noise samples and original samples into the neural network and compute the coverage value $value\_\mathcal{S}_{+noise}$. We then compare the difference between $value\_\mathcal{S}_{+adv}$ and $value\_\mathcal{S}_{+noise}$ and calculate the percentage increase of $value\_\mathcal{S}_{+adv}$ over $value\_\mathcal{S}_{+noise}$.
To further examine whether the observed coverage increase is truly related to errors, rather than resulting from statistical bias due to minimal impact, we perform repeated subsampling for each configuration (five runs) and conduct paired $t$-tests. The corresponding $p$-values are reported to assess the statistical significance of the differences.

Our objective is to assess the general error-revealing capability of neuron coverage criteria, rather than the behavior of specific adversarial attack operators. Accordingly, we generate erroneous samples using multiple representative adversarial attack methods and report coverage results by aggregating samples across different operators. This design allows us to focus on whether coverage criteria are more sensitive to prediction-changing perturbations than to label-preserving noise, while minimizing the influence of individual attack mechanisms.
Based on the two-phase experiment described above, we evaluate the sensitivity of the coverage criterion to adversarial samples by examining whether adding an adversarial sample set to the original samples increases the coverage. Additionally, by comparing the effects of adding noise samples versus adversarial samples to the original set, we aim to determine the primary factor contributing to the increase in coverage.

\input{table/table_rq11.tex}

\subsubsection{Results} \vspace{0.5em}\noindent

Table~\ref{tab:rq11} and Table~\ref{tab:rq11-gen} report the values of neuron coverage criteria for the original samples $\mathcal{S}$ and for the combined samples after adding adversarial examples $\mathcal{S}_{adv}$. Table~\ref{tab:rq11} corresponds to classification tasks, while Table~\ref{tab:rq11-gen} corresponds to generation tasks. 
Here, $DWB$, $TF$, and $TB$ denote adversarial attack methods for classification tasks, namely DeepWordBug, TextFooler, and TextBugger, respectively, while $MP$ and $SEQ$ represent attack methods for generation tasks, Morpheus and Seq2Sick.
Each major row, spanning one or more sub-rows, represents a specific coverage criterion. For classification tasks, the three main columns correspond to three representative transformer-based language models: Llama-2, BERT, and T5; for generation tasks, the models evaluated are Llama-2 and T5. Within each main column, four sub-columns are provided: the first shows the coverage value for the original samples $\mathcal{S}$, and the remaining three show the coverage values after adding adversarial samples generated by different attack methods ($\mathcal{S}+\mathcal{S}_{DWB}$, $\mathcal{S}+\mathcal{S}_{TF}$, and $\mathcal{S}+\mathcal{S}_{TB}$). For generation tasks, the corresponding adversarial attack methods are $\mathcal{S}+\mathcal{S}_{MP}$ and $\mathcal{S}+\mathcal{S}_{SEQ}$.
For each set of adversarial samples, the percentage increase in coverage value compared to the original samples is computed. This percentage increase is indicated by the transparency level of the gray background; a gray background indicates that the value is higher than that of the original samples, with darker shades representing greater increases.
For instance, in the first row, the initial coverage value, 0.966, corresponds to the NC with a threshold of 0.25 for the original sample \(\mathcal{S}\) under the Llama-2 model. The second value, 0.968, represents the coverage value after adding adversarial samples \(\mathcal{S}_{DWB}\) generated by the DeepWordBug attack under the same model and criterion, showing a 0.2\% increase; hence, a very light gray background is applied. Similarly, in the third row, the NC with a threshold of 0.75 yields a value of 0.371 for the original sample \(\mathcal{S}\), while the value for \(\mathcal{S} + \mathcal{S}_{DWB}\) is 0.417, indicating a 12.4\% increase, and thus a darker gray background is applied.

As shown in Table~\ref{tab:rq11}, it can be observed that the majority of the background cells are colored, indicating that for a given original sample \(S\), the addition of adversarial samples \(S_{\text{adv}}\) generally leads to an increase in coverage across nearly all criteria. For the NC criterion, when the threshold is set at 0.25 or 0.50, the coverage for the original sample alone is already close to 1 in some cases. For instance, the BERT model achieves 100\% and 99.9\% coverage, respectively, so adding adversarial samples does not result in further improvement. However, when the threshold is set at 0.75, the coverage value increases across an entire row.
The KMNC criterion behaves similarly to NC. As the threshold increases from 10 to 100, the granularity of coverage becomes finer, and the background color darkens, indicating a larger percentage increase. This occurs because finer granularity divides the coverage space into smaller intervals, making it easier to discover new coverage areas, thereby allowing more room for improvement. The criteria NBC and SNAC, which detect neurons that exceed the upper bound or fall below the lower bound, and CC, which detects deviations from normal causal logic, all play crucial roles in identifying outliers. Consequently, the addition of adversarial samples results in a significant increase in coverage for these criteria.
For the NLC criterion, adding adversarial samples to the original samples does not improve coverage in the original version of NLC under any conditions. In contrast, the NLC+ version shows improvement in some cases. This discrepancy arises because the original NLC is based on overall variance; adding samples similar to existing ones in the set may reduce overall variance. 
Specifically, if the difference between the new sample \(x_{n+1}\) and the original mean \(\mu\) satisfies the following condition:
$|x_{n+1} - \mu| < \sqrt{\sigma^2 \cdot n / ( n + 1 )}$,
where \(\sigma^2\) is the original variance, \(n\) is the number of original samples, and \(|x_{n+1} - \mu|\) is the distance between the new sample and the original mean.
Although NLC+ attempts to address this by ignoring samples that reduce variance during incremental updates, this introduces a new issue: NLC+ becomes highly sensitive to the order of samples. Different sequences of samples can yield different results, which is why NLC+ only improves coverage in certain situations.

\input{table/table_rq12}

Table ~\ref{tab:rq12} presents the changes in coverage criteria after adding the noise sample set $\mathcal{S}_{\text{noise}}$ and the adversarial sample set $\mathcal{S}_{\text{adv}}$ to the original sample $\mathcal{S}$ across three classic transformer architectures (with each value representing an average for that architecture). Each major row, spanning two sub-rows, represents a specific transformer structure, where "DEC.", "ENC.", and "D.$+$E." represent the average results for the decoder, encoder, and encoder-decoder models, respectively. 
In the two sub-rows, $\delta_{\text{noise}}$ indicates the change resulting from adding the noise sample set $\mathcal{S}_{\text{noise}}$, while $\delta_{\text{adv}}$ represents the change from adding the adversarial sample set $\mathcal{S}_{\text{adv}}$. When $\delta_{\text{adv}}$ exceeds $\delta_{\text{noise}}$, the corresponding value is bolded. 
For example, in the first column, the first two rows both show 0.002, indicating that the increase in NC with a threshold of 0.25 is 0.002 for both sample sets; thus, no bolding is applied. In contrast, in the second column, the values 0.004 and 0.009 indicate that the increase in NC with a threshold of 0.50 is greater when adding $\mathcal{S}_{\text{adv}}$, so the latter value is bolded.

It can be observed that in most cases, the values corresponding to $\delta_{\text{adv}}$ are bolded, indicating that the increase in coverage is greater after the addition of the adversarial sample set than after adding the noise sample set. For the NC and KMNC criteria, it is apparent that in cases where the values are not bolded, the increments are often identical, particularly when the thresholds are lower. As discussed earlier, lower thresholds result in coarser coverage, making it easier to miss finer changes, which explains why the increments in coverage from the two different sample sets are sometimes exactly the same.
For NLC and NLC+, some cases are bolded while others are not. As previously mentioned, this is because the variance- and covariance-based nature of these criteria can lead to variations that decrease the values when certain samples are added. Moreover, the order in which the samples are added also plays a role, leading to different outcomes.
The total increase in coverage value after adding the noise sample set is 2.757, while the total increase after adding the adversarial sample set is 2.849, representing a minimal average improvement of 0.28\% per case.
In addition, to further assess whether these minimal improvements are reliable, Table~\ref{tab:rq12_pvalues} reports the $p$-values obtained from paired $t$-tests across all configurations. Improvements with $p < 0.05$ are considered statistically significant. Specifically, the cells with a gray background and dashes ``--'' correspond to cases where the coverage value of the original samples is already saturated, resulting in zero increment. Among the remaining cases, cells highlighted in green indicate $p < 0.05$, while those in red correspond to $p \ge 0.05$. As shown, the NLC and NLC$+$ criteria consistently yield $p \ge 0.05$ across all settings, reflecting their sensitivity to input order. Most of the other red cells are concentrated in NC and KMNC at thresholds of 0.5 and 50, respectively, where the initial coverage is already high and thus the incremental gains are negligible, leading to weak significance. In addition, although NBC under the Encoder--Decoder configuration is marked red, its $p$-value is only slightly above 0.05.

In summary, except for the NLC/NLC+ criteria, which focus on intra-layer neuron combinations, the factors contributing to coverage increase generally encapsulate the semantic errors introduced by adversarial samples. However, this additional semantic information accounts for only approximately a $0.32\%$  improvement.
Under the current computation paradigm of coverage criteria, most of the uncovered regions can already be explored through random perturbations, while only a very small fraction truly requires error-related information to be activated.

\begin{tcolorbox}[title=Finding 1:,colbacktitle=gray!50, coltitle=black,fonttitle=\bfseries,boxrule=0.8pt]
	{ \textbf{Although the addition of error samples enhances the neuron coverage of a test suite (except for NLC/NLC+), the resulting coverage increment over noise samples is minimal, at just 0.32\%, reflecting that the increment mainly arises from adding new samples, with little extra increase due to the error semantics of the samples.
	} }
\end{tcolorbox}

\begin{tcolorbox}[title=Implication 1:,colbacktitle=gray!50, coltitle=black,fonttitle=\bfseries,boxrule=0.8pt]
	{ 
	\begin{compactitem}[$\bullet$]
		\item  \textbf{Usage.} The coverage gains from adversarial samples over benign noise are marginal and offer limited practical value for reliable error detection in test suites.
		
		\item \textbf{Design.} Future criteria should distinguish erroneous from normal samples, amplifying error-related semantics while suppressing noise-like perturbations in adversarial inputs.
	\end{compactitem}
	 }
\end{tcolorbox}

%% file: figure/figure8.tex
\begin{figure}[t]
	\centering
	\includegraphics[scale=0.65]{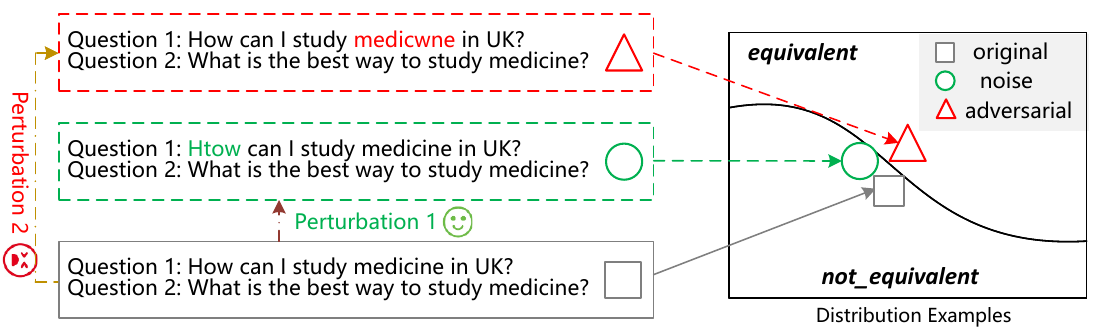}
	\caption{Distribution Examples of Different Samples. Given a standard sample (represented by squares), perturbations (noise) can be added to generate new samples. Perturbations that alter the semantic meaning of the sample are termed adversarial samples (represented by triangles), while those that do not are classified as regular noise samples (represented by circles).}
	\label{fig:RQ1}
 \vspace{-5pt}
\end{figure}

%% file: table/table_rq11.tex
\begin{table*}[]
	\centering
	\caption{The average coverage values achieved by the original samples $\mathcal{S}$ and various error-revealing (text) test suites $\mathcal{S}_{adv}$ on transformer-based language models for classification tasks. The cells corresponding to $\mathcal{S} + \mathcal{S}_{adv}$ are shaded in \colorbox{gray}{\tiny{\textcolor{white}{\raisebox{0.5ex}{gray}}}}, with darker shades indicating a higher percentage increase in coverage over $\mathcal{S}$. Cells without shading indicate no increase.}
	\label{tab:rq11}
	\resizebox{.95\textwidth}{!}{
		\begin{tabular}{ll|cccccccccccc@{}}
			\toprule
			\multirow{2}{*}{\makecell[c]{\textbf{Coverage} \\ \textbf{Criteria}}} &  {\multirow{2}{*}{\textbf{Config.}}} & \multicolumn{4}{c}{\textbf{LLama-2}} & \multicolumn{4}{c}{\textbf{BERT}} & \multicolumn{4}{c}{\textbf{T5}} \\ 
			
			& & $\mathcal{S}$ & {$\mathcal{S}+\mathcal{S}_{DWB}$} & {$\mathcal{S} + \mathcal{S}_{TF}$} & {$\mathcal{S} + \mathcal{S}_{TB}$} & $\mathcal{S}$ & {$\mathcal{S}+\mathcal{S}_{DWB}$} & {$\mathcal{S} + \mathcal{S}_{TF}$} & {$\mathcal{S} + \mathcal{S}_{TB}$} & $\mathcal{S}$ & {$\mathcal{S}+\mathcal{S}_{DWB}$} & {$\mathcal{S} + \mathcal{S}_{TF}$} & {$\mathcal{S} + \mathcal{S}_{TB}$} \\ \midrule
			{\multirow{3}{*}{\textbf{NC}}} & {T=0.25} & 0.966 & \colorbox{gray!10}{0.968} & \colorbox{gray!10}{0.967} & \colorbox{gray!10}{0.968} & 1.000 & 1.000 & 1.000 & 1.000 & 0.991 & \colorbox{gray!10}{0.992} & \colorbox{gray!10}{0.992} & \colorbox{gray!10}{0.992} \\
			& {T=0.50} & 0.603 & \colorbox{gray!13}{0.616} & \colorbox{gray!10}{0.606} & \colorbox{gray!10}{0.608} & 0.999 & 0.999 & 0.999 & 0.999 & 0.970 & \colorbox{gray!11}{0.976} & \colorbox{gray!11}{0.975} & \colorbox{gray!10}{0.972} \\
			& {T=0.75} & 0.371 & \colorbox{gray!28}{0.417} & \colorbox{gray!25}{0.401} & \colorbox{gray!26}{0.407} & 0.830 & \colorbox{gray!23}{0.857} & \colorbox{gray!22}{0.852} & \colorbox{gray!24}{0.864} & 0.701 & \colorbox{gray!24}{0.732} & \colorbox{gray!23}{0.728} & \colorbox{gray!23}{0.727} \\ \midrule
			{\multirow{3}{*}{\textbf{KMNC}}} & {K=10} & 0.989 & \colorbox{gray!10}{0.995} & \colorbox{gray!10}{0.993} & \colorbox{gray!10}{0.993} & 0.996 & \colorbox{gray!10}{0.998} & \colorbox{gray!10}{0.998} & \colorbox{gray!10}{0.998} & 0.987 & \colorbox{gray!07}{0.994} & \colorbox{gray!06}{0.993} & \colorbox{gray!05}{0.992} \\
			& {K=50} & 0.923 & \colorbox{gray!23}{0.956} & \colorbox{gray!22}{0.946} & \colorbox{gray!22}{0.950} & 0.960 & \colorbox{gray!20}{0.979} & \colorbox{gray!16}{0.975} & \colorbox{gray!20}{0.979} & 0.921 & \colorbox{gray!15}{0.953} & \colorbox{gray!15}{0.947} & \colorbox{gray!13}{0.946} \\
			& {K=100} & 0.875 & \colorbox{gray!35}{0.921} & \colorbox{gray!33}{0.906} & \colorbox{gray!34}{0.913} & 0.930 & \colorbox{gray!33}{0.960} & \colorbox{gray!32}{0.954} & \colorbox{gray!33}{0.960} & 0.873 & \colorbox{gray!35}{0.918} & \colorbox{gray!29}{0.909} & \colorbox{gray!29}{0.909} \\ \midrule
			\textbf{NBC} & {None} & 0.450 & \colorbox{gray!50}{0.675} & \colorbox{gray!37}{0.616} & \colorbox{gray!45}{0.655} & 0.496 & \colorbox{gray!41}{0.701} & \colorbox{gray!33}{0.657} & \colorbox{gray!41}{0.704} & 0.497	 & \colorbox{gray!41}{0.693} & \colorbox{gray!35}{0.663} & \colorbox{gray!40}{0.674} \\ \midrule
			\textbf{SNAC} & {None} & 0.451 & \colorbox{gray!51}{0.680} & \colorbox{gray!37}{0.620} & \colorbox{gray!46}{0.658} & 0.499 & \colorbox{gray!41}{0.701} & \colorbox{gray!32}{0.659} & \colorbox{gray!41}{0.704} & 0.498 & \colorbox{gray!40}{0.702} & \colorbox{gray!37}{0.667} & \colorbox{gray!36}{0.682} \\ \midrule
			\textbf{NLC} & \multirow{2}{*}{None} & 0.180 & 0.177 & 0.169 & 0.162 & 0.193 & 0.182 & 0.176 & 0.185 & 0.188 & 0.154 & 0.167 & 0.160 \\
			\textbf{NLC+} &  & 0.660 & \colorbox{gray!18}{0.675} &  {0.654} & \colorbox{gray!25}{0.702} & 0.672 & {0.655} & \colorbox{gray!28}{0.694} & \colorbox{gray!32}{0.707} & 0.479 & \colorbox{gray!10}{0.480} & \colorbox{gray!18}{0.498} & \colorbox{gray!15}{0.495} \\	
			\midrule		
			$\textbf{CC}_{w}$ & {None} & 0.017 & \colorbox{gray!35}{0.023} & \colorbox{gray!47}{0.025} & \colorbox{gray!41}{0.024} & 0.027 & \colorbox{gray!41}{0.038} & \colorbox{gray!44}{0.039} & \colorbox{gray!30}{0.035} & 0.019 & \colorbox{gray!26}{0.024} & \colorbox{gray!42}{0.027} & \colorbox{gray!36}{0.026} \\

			 \bottomrule
		\end{tabular}
		}
\vspace{-10pt}
\end{table*}

\begin{table*}[]
	\centering
	\caption{The average coverage values achieved by the original samples $\mathcal{S}$ and two error-revealing (text) test suites $\mathcal{S}_{adv}$ on transformer-based language models for  generation tasks. The cells corresponding to $\mathcal{S} + \mathcal{S}_{adv}$ are shaded in \colorbox{gray}{\tiny{\textcolor{white}{\raisebox{0.5ex}{gray}}}}, with darker shades indicating a higher percentage increase in coverage over $\mathcal{S}$. Cells without shading indicate no increase.}
	\label{tab:rq11-gen}
	\resizebox{.55\textwidth}{!}{
		\begin{tabular}{ll|cccccc@{}}
			\toprule
			\multirow{2}{*}{\makecell[c]{\textbf{Coverage} \\ \textbf{Criteria}}} &  {\multirow{2}{*}{\textbf{Config.}}} 
			& \multicolumn{3}{c}{\textbf{LLama-2}} & \multicolumn{3}{c}{\textbf{T5}} \\ 
			& & $\mathcal{S}$ & {$\mathcal{S} + \mathcal{S}_{MP}$} & {$\mathcal{S} + \mathcal{S}_{SEQ}$} 
			& $\mathcal{S}$ & {$\mathcal{S} + \mathcal{S}_{MP}$} & {$\mathcal{S} + \mathcal{S}_{SEQ}$} \\ 
			\midrule
			{\multirow{3}{*}{\textbf{NC}}} 
			& {T=0.25} & 0.892 & \colorbox{gray!10}{0.898} & \colorbox{gray!11}{0.902} & 0.912 & \colorbox{gray!10}{0.917} & \colorbox{gray!10}{0.916} \\
			& {T=0.50} & 0.542 & \colorbox{gray!10}{0.546} & \colorbox{gray!10}{0.548} & 0.878 & \colorbox{gray!11}{0.887} & \colorbox{gray!10}{0.884} \\
			& {T=0.75} & 0.310 & \colorbox{gray!25}{0.329} & \colorbox{gray!26}{0.333} & 0.609 & \colorbox{gray!23}{0.622} & \colorbox{gray!23}{0.621} \\ 
			\midrule
			{\multirow{3}{*}{\textbf{KMNC}}} 
			& {K=10}  & 0.955 & \colorbox{gray!10}{0.961} & \colorbox{gray!10}{0.960} & 0.944 & \colorbox{gray!07}{0.950} & \colorbox{gray!06}{0.949} \\
			& {K=50}  & 0.846 & \colorbox{gray!22}{0.866} & \colorbox{gray!22}{0.869} & 0.829 & \colorbox{gray!15}{0.849} & \colorbox{gray!14}{0.846} \\
			& {K=100} & 0.792 & \colorbox{gray!33}{0.814} & \colorbox{gray!34}{0.820} & 0.780 & \colorbox{gray!29}{0.802} & \colorbox{gray!29}{0.801} \\ 
			\midrule
			\textbf{NBC} & {None} & 0.388 & \colorbox{gray!36}{0.515} & \colorbox{gray!42}{0.544} & 0.427 & \colorbox{gray!35}{0.572} & \colorbox{gray!40}{0.583} \\ 
			\midrule
			\textbf{SNAC} & {None} & 0.389 & \colorbox{gray!37}{0.517} & \colorbox{gray!43}{0.545} & 0.428 & \colorbox{gray!37}{0.576} & \colorbox{gray!36}{0.588} \\ 
			\midrule
			\textbf{NLC} & \multirow{2}{*}{None} & 0.142 & 0.133 & 0.128 & 0.146 & 0.133 & 0.127 \\
			\textbf{NLC+} &  & 0.582 & 0.573 & \colorbox{gray!25}{0.606} & 0.412 & 0.410 & 0.415 \\ 
			\midrule
			$\textbf{CC}_{w}$ & {None} & 0.013 & \colorbox{gray!47}{0.019} & \colorbox{gray!41}{0.018} & 0.009& \colorbox{gray!42}{0.012} & \colorbox{gray!36}{0.015} \\
			\bottomrule
		\end{tabular}
	}
	\vspace{-10pt}
\end{table*}

%% file: table/table_rq12.tex
\begin{table*}[]
	\centering
	\caption{The change in coverage value after adding the noise sample  $\mathcal{S}_{noise}$ and the adversarial sample  $\mathcal{S}_{adv}$ to the original sample $\mathcal{S}$. The cells with \textbf{bolded} values indicate that the coverage increase from adding $\mathcal{S}_{adv}$ is greater than that from adding $\mathcal{S}_{noise}$. Here, \textbf{DEC.}, \textbf{ENC.}, and \textbf{D. $+$ E.} denote \textit{decoder-only}, \textit{encoder-only}, and \textit{encoder–decoder} models, respectively.
	}
	\label{tab:rq12}
	\resizebox{.95\textwidth}{!}{
		\begin{tabular}{llccccccccccc@{}}
			\toprule
			\multicolumn{2}{c}{\multirow{2}{*}{\makecell[c]{\textbf{Configuration}}}} & \multicolumn{3}{c}{\textbf{NC}} & \multicolumn{3}{c}{\textbf{KMNC}} & \multirow{2}{*}{\makecell[c]{\textbf{NBC}}} & \multirow{2}{*}{\makecell[c]{\textbf{SNAC}}} & \multirow{2}{*}{\makecell[c]{\textbf{NLC}}} & \multirow{2}{*}{\makecell[c]{\textbf{NLC$+$}}} & \multirow{2}{*}{\makecell[c]{$\textbf{CC}_w$}} \\ 
			
			& & T=0.25 & T=0.50 & T=0.75 & K=10 & K=50 & K=100 & & & & & \\ \midrule
			\multirow{2}{*}{\textbf{DEC.}} & $\delta_{{noise}}$ & 0.002 & 0.004 & 0.019 & 0.004 & 0.026 & 0.038 & 0.216 & 0.218 & -0.019 & 0.025 & 0.001 \\
			& $\delta_{adv}$ & 0.002 & \textbf{0.009} & \textbf{0.041} & \textbf{0.005} & \textbf{0.030} & \textbf{0.042} & \textbf{0.221} & \textbf{0.226} & \textbf{-0.013} & \textbf{0.018} & \textbf{0.006} \\ \midrule
			
			\multirow{2}{*}{\textbf{ENC.}} & $\delta_{{noise}}$ & 0.000 & 0.000 & 0.024 & 0.002 & 0.018 & 0.028 & 0.193 & 0.194 & -0.007 & 0.016 & 0.007 \\
			& $\delta_{adv}$ & 0.000 & 0.000 & \textbf{0.026} & 0.002 & 0.018 & \textbf{0.029} & \textbf{0.198} & \textbf{0.201} & -0.013 & 0.014 & \textbf{0.010} \\ \midrule
			
			\multirow{2}{*}{\textbf{D. $+$ E.}} & $\delta_{{noise}}$ &0.000 & 0.014 & 0.028 & 0.006 & 0.030 & 0.042 & 0.208 & 0.206 & -0.033 & 0.005 & 0.005 \\
			& $\delta_{adv}$ & \textbf{0.001} & 0.014 & \textbf{0.030} & \textbf{0.007} & 0.030 & \textbf{0.044} & 0.199 & \textbf{0.209} & \textbf{-0.028} & 0.012 & \textbf{0.006} \\
			\bottomrule
		\end{tabular}
	}
\vspace{-11pt}
\end{table*}

\begin{table*}[]
	\centering
\caption{Paired $t$-test ( $\delta_{adv} > \delta_{noise}$) p-values based on 5  subsampling runs under each configuration. Cells with $p<0.05$ are highlighted in light green, those with $p\ge 0.05$ in light red, and cases with no change in coverage are denoted by a dash ``--''.}

	\label{tab:rq12_pvalues}
	\resizebox{.95\textwidth}{!}{
		\begin{tabular}{lccccccccccc@{}}
			\toprule
			\multirow{2}{*}{\makecell[c]{\textbf{Configuration}}} 
			& \multicolumn{3}{c}{\textbf{NC}} 
			& \multicolumn{3}{c}{\textbf{KMNC}} 
			& \multirow{2}{*}{\makecell[c]{\textbf{NBC}}} 
			& \multirow{2}{*}{\makecell[c]{\textbf{SNAC}}} 
			& \multirow{2}{*}{\makecell[c]{\textbf{NLC}}} 
			& \multirow{2}{*}{\makecell[c]{\textbf{NLC$+$}}} 
			& \multirow{2}{*}{\makecell[c]{$\textbf{CC}_{w}$}} \\
			& T=0.25 & T=0.50 & T=0.75 & K=10 & K=50 & K=100 & & & & & \\ 
			\midrule
			
			\textbf{DEC.} 
			& \cellcolor{gray!15}-- 
			& \cellcolor{green!15}0.014 
			& \cellcolor{green!15}0.002 
			&  \cellcolor{gray!15}-- 
			& \cellcolor{green!15}0.017 
			& \cellcolor{green!15}0.005 
			& \cellcolor{green!15}0.020 
			& \cellcolor{green!15}0.019 
			& \cellcolor{red!15}0.248 
			& \cellcolor{red!15}0.823 
			& \cellcolor{green!15}0.004 \\
			
			\textbf{ENC.} 
			&  \cellcolor{gray!15}-- 
			& \cellcolor{gray!15} -- 
			& \cellcolor{green!15}0.006 
			&  \cellcolor{gray!15}-- 
			& \cellcolor{red!15}0.187 
			& \cellcolor{green!15}0.031 
			& \cellcolor{green!15}0.007 
			& \cellcolor{green!15}0.003 
			& \cellcolor{red!15}0.315 
			& \cellcolor{red!15}0.588 
			& \cellcolor{green!15}0.008 \\
			
			\textbf{D. $+$ E.} 
			&  \cellcolor{gray!15}-- 
			& \cellcolor{red!15}0.500 
			& \cellcolor{green!15}0.035
			&  \cellcolor{gray!15}-- 
			& \cellcolor{red!15}0.648 
			& \cellcolor{green!15}0.016 
			& \cellcolor{red!15}0.064 
			& \cellcolor{green!15}0.021 
			& \cellcolor{red!15}0.622 
			& \cellcolor{red!15}0.622 
			& \cellcolor{green!15}0.003 \\
			\bottomrule
		\end{tabular}
	}
	\vspace{-10pt}
\end{table*}

%% file: chapter/evaluation2.tex
\subsection{Distribution Diversity ( RQ2 )}
	
Distribution diversity refers to the dispersion of a test suite across the sample space.
This research question aims to explore whether neuron coverage criteria can effectively evaluate the distribution diversity of a test suite. Specifically, it seeks to determine if, given test suites of equal size, coverage criteria can identify a test suite with higher dispersion.

\subsubsection{Experiment Design} \vspace{0.5em}\noindent

\input{figure/figure3-2.tex}
\input{figure/figure3-22.tex}
To address RQ2, we will design test suites with varying degrees of distribution diversity and analyze whether the coverage values generated by these test suites when passed through the model are positively correlated with their distribution diversity. Figure ~\ref{figure3-2} provides an intuitive illustration of distribution diversity. Specifically, in the context of a sentiment classification task, circles and squares represent two different test suites. Each test suite contains blue, yellow, and red elements, corresponding to samples from the positive, negative, and neutral categories, respectively. The distribution ranges for these categories are outlined with dashed lines.
In terms of overall distribution, the test suite represented by circles has samples evenly spread across the entire sample space, indicating high diversity. In contrast, the test suite represented by squares has samples that are more concentrated in specific areas of the sample space; for instance, four closely clustered samples appear in the upper-right corner of the positive region. Specific sample examples are provided on the right side of the figure: A and B are two highly similar, clustered samples that differ only by the adverbs ``really'' and ``truly'' (as underlined), while C is a completely distinct positive sample with different semantics.
Following the design in~\cite{nlc}, given the original sample set $\mathcal{S}$, as shown in Figure~\ref{figure3-22} we construct a perturbed subsample set $\mathcal{PSS}$ with the same size as $\mathcal{S}$. This ensures equal sample quantity while reducing distributional diversity. To further amplify the diversity factor, we expand $\mathcal{PSS}$ to obtain $\mathcal{PSS}^+$, which contains more samples than $\mathcal{S}$ but exhibits even lower diversity. This design highlights the sensitivity of the coverage criteria to variations in diversity.

\noindent \textit{Perturbed Subsample Set (denoted as \(\mathcal{PSS}\)):}
\[
\mathcal{PSS} = \mathcal{S}_{\text{sub}} \cup \left\{ \tilde{x}_i^j = \text{Perturb}(x_i) \mid j = 1, 2, \dots, k \right\}; \quad \sum_{i=1}^{|\mathcal{S}_{\text{sub}}|} k = |\mathcal{S}| - |\mathcal{S}_{\text{sub}}|, \left| \mathcal{PSS} \right| = \left| \mathcal{S} \right|.
\]

This formula represents a sample set that is generated by perturbing a subset of the original sample \(\mathcal{S}\) to match the size of the original sample. Specifically, \(\mathcal{S}_{\text{sub}}\) refers to a small subset of the original sample, comprising 10\% of \(\mathcal{S}\). The perturbation function \(\text{Perturb}(\cdot)\) applies minor random perturbations to each sample \(x_i \in \mathcal{S}_{\text{sub}}\) \(k\) times, where \(k\) is chosen such that the total number of perturbed samples equals the size of the original sample: \(\sum_{i=1}^{|\mathcal{S}_{\text{sub}}|} k = |\mathcal{S}| - |\mathcal{S}_{\text{sub}}|\). Thus, the total size of \(\mathcal{PSS}\) equals that of the original sample set \(\mathcal{S}\).

\noindent \textit{Perturbed Extended Subsample Set (denoted as \(\mathcal{PSS_{+}}\)):}

\[
\mathcal{PSS+} = \mathcal{S}_{\text{sub}} \cup \left\{ \tilde{x}_i^j = \text{Perturb}(x_i) \mid j = 1, 2, \dots, k \right\}; \quad \sum_{i=1}^{|\mathcal{S}_{\text{sub}}|} k = m \times |\mathcal{S}| - |\mathcal{S}_{\text{sub}}|,  \left| \mathcal{PSS+} \right| = \left| \mathcal{S} \right|.
\]

This formula describes a sample set generated by perturbing a subset of the original sample $\mathcal{S}$ to a size that is \(m\) times the size of the original sample. The perturbation process is identical to that used for \(\mathcal{PSS}\), with the only difference being that the value of \(k\) is chosen such that the size of \(\mathcal{PSS_{+}}\) is \(m\) times the size of the original sample: \(\sum_{i=1}^{|\mathcal{S}_{\text{sub}}|} k =  m \times |\mathcal{S}| - |\mathcal{S}_{\text{sub}}|\).

Having established the three test suites, the diversity of these test suites can be ranked in descending order as \(\mathcal{S} > \mathcal{PSS_{+}} > \mathcal{PSS}\). This ranking is justified by the composition of the original sample set, which includes 5000 diverse text samples from real-world datasets. In contrast, \(\mathcal{PSS}\) is based on perturbations of only 100 text samples, leading to a concentrated distribution around these 100 samples despite matching the original sample size. Although \(\mathcal{PSS_{+}}\) has a distribution that is more diverse than \(\mathcal{PSS}\) due to its larger size, its diversity is still significantly less than that of the original sample set because it remains centred around the 100 samples, albeit with a higher quantity.
Therefore, we construct the three aforementioned test suites and introduce them into transformer-based language models, then calculate the coverage value for each criterion. Referring to the SOTA NLC evaluation method ~\cite{nlc} for coverage criteria, we establish the ground truth for the coverage values induced by these test suites as \(\mathcal{C}_{\mathcal{S}} > \mathcal{C}_{\mathcal{PSS_{+}}} > \mathcal{C}_{\mathcal{PSS}}\). Finally, we examine the extent to which the neuron coverage criterion aligns with this ground truth under various configurations.

\subsubsection{Results} \vspace{0.5em}\noindent

\input{table/table_rq2.tex}

Table~\ref{tab:rq2} and Table~\ref{tab:rq2-gen} present the average coverage values obtained by the sample sets \(\mathcal{S}\), \(\mathcal{PSS_{+}}\), and \(\mathcal{PSS}\).
Specifically, Table~\ref{tab:rq2} reports results for classification tasks, while Table~\ref{tab:rq2-gen} reports results for generation tasks.
Each major row corresponds to a specific coverage criterion, while the main columns represent the models, further subdivided into three sub-columns indicating the respective sample sets. The table is designed to assess whether the coverage values align with the expected ground truth relationship \(\mathcal{C}_{\mathcal{S}} > \mathcal{C}_{\mathcal{PSS_{+}}} > \mathcal{C}_{\mathcal{PSS}}\), with a checkmark indicating when this condition is met. The final column summarizes the number of instances where this ground truth relationship holds.
For example, in the first row, the coverage values \(0.966\), \(0.955\), and \(0.949\) correspond to Neuron Coverage (NC) with a threshold of 0.25 under the LLama-2 model. Since \(0.966 > 0.955 > 0.949\) satisfies the expected relationship, a checkmark is placed accordingly.

As shown in Table~\ref{tab:rq2}, for NC across thresholds \(T=0.25\), \(T=0.50\), and \(T=0.75\), the relationship \(\mathcal{C}_{\mathcal{S}} > \mathcal{C}_{\mathcal{PSS_{+}}} > \mathcal{C}_{\mathcal{PSS}}\) is generally maintained, particularly at higher thresholds. All three models conform to this relationship at \(T=0.75\), with 6 out of 9 configurations meeting the expected pattern. This is partly because, at lower thresholds such as 0.25 or 0.50, the coverage for the original sample set \(\mathcal{S}\) often approaches saturation, rendering it difficult to differentiate the coverage values across the three sample sets.
For the K-Multisection Neuron Coverage (KMNC) criterion, a strong alignment with the expected relationship is observed, particularly at \(K=10\) and \(K=100\), where all models adhere to the relationship. However, at \(K=50\), the LLama-2 model deviates from this pattern, slightly reducing the overall satisfaction rate to 8 out of 9 configurations.
The Neuron Boundary Coverage (NBC) and Strong Neuron Activation Coverage (SNAC) criteria consistently satisfy the expected relationship across all configurations and models. This consistency underscores the robustness of these criteria in distinguishing between different sample sets, with all configurations demonstrating the expected pattern, thereby validating their effectiveness in assessing distribution diversity.
In contrast, the Neuron Linear Coverage (NLC) criterion exhibits more variability. For standard NLC, the expected relationship holds in one out of three configurations, whereas the enhanced version, NLC+, satisfies the relationship in two out of three cases, resulting in an overall mixed outcome of 3 out of 6 configurations. This variability aligns with prior analyses, indicating that the sensitivity of this metric can lead to more random results.
Lastly, the Composite Coverage (CC) criterion consistently adheres to the expected relationship across all models and configurations. The uniformity across configurations suggests that CC is particularly sensitive to differences between sample sets, with even slight perturbations in noise samples rarely affecting the overall CC coverage.
Table~\ref{tab:rq2-gen} shows similar findings for generation tasks, with the ground-truth relationship satisfied in all cases except for the NLC series and NC at the low threshold  \(T=0.25\).
Moreover, to further demonstrate the correlation between coverage and distributional diversity, Figure~\ref{figure_rq2} presents one representative configuration for both classification and generation tasks. Under a fixed test suite size, diversity is gradually increased by perturbing larger sub-samples. The results indicate a generally positive correlation, as coverage consistently increases with higher diversity.

\input{figure/figure_rq2}
In conclusion, the table demonstrates that the relationship \(\mathcal{C}_{\mathcal{S}} > \mathcal{C}_{\mathcal{PSS_{+}}} > \mathcal{C}_{\mathcal{PSS}}\) is  largely maintained across different models and criteria. NBC, SNAC, and CC criteria show the most consistent adherence to this pattern, indicating their effectiveness in distinguishing sample sets and assessing coverage improvements. Although NC and KMNC may struggle to capture subtle variations at lower thresholds, they generally follow the expected relationship. Conversely, the variability of the NLC criterion suggests it may be less reliable under certain conditions or for specific models.

\begin{tcolorbox}[title=Finding 2:,colbacktitle=gray!50, coltitle=black,fonttitle=\bfseries,boxrule=0.8pt]
	{ \textbf{Neuron coverage criteria effectively reflect the distribution diversity of a test suite in 44 out of 56 cases, though NLC shows some randomness. For threshold-based criteria, higher thresholds better capture subtle distribution variations.
	} }
\end{tcolorbox}

\begin{tcolorbox}[title=Implication 2:,colbacktitle=gray!50, coltitle=black,fonttitle=\bfseries,boxrule=0.8pt]
	{ 
		\begin{compactitem}[$\bullet$]
		\item \textbf{Usage.} Existing criteria (except the  NLC series) can be applied to evaluate the distributional diversity of test suites. However, criteria with thresholds require manual tuning, since overly small thresholds may fail to capture meaningful differences. 
		\item  \textbf{Design.} Future criteria should be designed to remain robust against variations in input order while preserving their ability to differentiate distributional diversity.

		\end{compactitem}
	}
\end{tcolorbox}

%% file: figure/figure3-2.tex
\begin{figure}[t]
	\centering
	\includegraphics[scale=0.16]{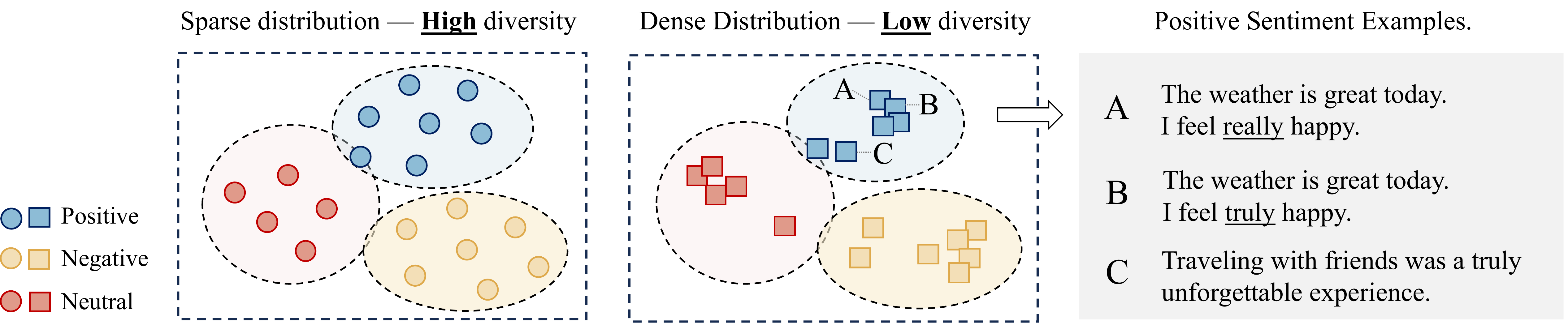}

	\caption{An illustration of distribution diversity in sentiment classification, comparing test suites of equal size. The first sub-figure shows high diversity with a widely dispersed test suite, while the second depicts low diversity with a tightly clustered test suite.}
	\label{figure3-2}
 \vspace{-5pt}
\end{figure}

%% file: figure/figure3-22.tex
\begin{figure}[t]
	\centering
	\includegraphics[scale=0.65]{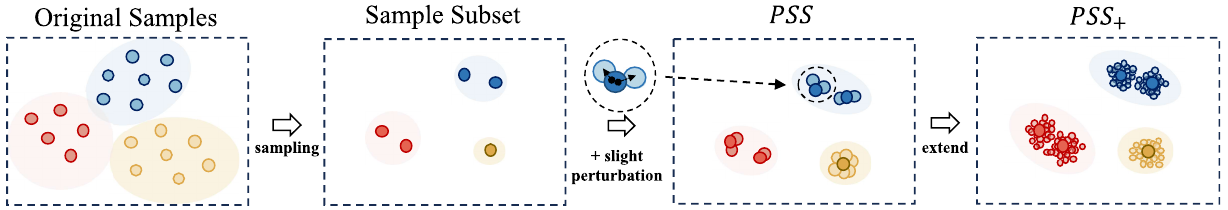}

	\caption{Process for constructing the \(\mathcal{PSS}\) and \(\mathcal{PSS_{+}}\) test suites. A random subset of samples is first selected from the original samples, then slight perturbations are applied to expand this subset to the original sample size, yielding \(\mathcal{PSS}\) and \(\mathcal{PSS_{+}}\).}
	\label{figure3-22}
\vspace{-5pt}
\end{figure}

%% file: table/table_rq2.tex
\begin{table*}[]
	\centering
	\caption{The average coverage values achieved by the sample sets \({\mathcal{S}}\), \({\mathcal{PSS_{+}}}\), and \({\mathcal{PSS}}\) on transformer-based language models for classification tasks. A checkmark indicates that the relationship \(\mathcal{C}_{\mathcal{S}} > \mathcal{C}_{\mathcal{PSS_{+}}} > \mathcal{C}_{\mathcal{PSS}}\) is satisfied. The final column shows the number of cases where this ground truth relationship holds.}
	\label{tab:rq2}
	\resizebox{.95\textwidth}{!}{
		\begin{tabular}{ll|cccc cccc cccc c c@{}}
			\toprule
			\multirow{2}{*}{\makecell[c]{\textbf{Coverage} \\ \textbf{Criteria}}} &  {\multirow{2}{*}{\textbf{Config.}}} &
			\multicolumn{4}{c}{ \textbf{LLama-2}} & \multicolumn{4}{c}{\textbf{BERT}} & \multicolumn{4}{c}{\textbf{T5}} & \multicolumn{2}{c}{\multirow{2}{*}{\makecell[c]{\textbf{Groundtruth} \\  \textbf{Satisfaction}}}} \\ 
			&   & $\mathcal{S}$    & $\mathcal{PSS_{+}}$ & $\mathcal{PSS}$ &   & $\mathcal{S}$    & $\mathcal{PSS_{+}}$ & $\mathcal{PSS}$ &   & $\mathcal{S}$    & $\mathcal{PSS_{+}}$ & $\mathcal{PSS}$ &   &  &  \\ \midrule
			\multirow{3}{*}{NC}       & T=0.25                   & 0.966 & 0.955 & 0.949  & \checkmark & 1.000 & 1.000 & 1.000 &  & 0.991 & 0.986 & 0.991  &  & 1/3 & \multirow{3}{*}{6/9} \\
			& T=0.50                   & 0.603 & 0.588 & 0.586     & \checkmark & 0.999 & 0.999 & 0.999 & & 0.970 & 0.963 & 0.958  & \checkmark & 2/3 &  \\
			& T=0.75                   & 0.371 & 0.313 & 0.284  & \checkmark & 0.830 & 0.746 & 0.657 & \checkmark & 0.701 & 0.687 & 0.672  & \checkmark & 3/3 &  \\
			\midrule
			\multirow{3}{*}{KMNC}     & K=10                     & 0.989 & 0.947 & 0.922  & \checkmark & 0.996 & 0.968 & 0.950 & \checkmark & 0.987 & 0.924 & 0.899  & \checkmark & 3/3 & \multirow{3}{*}{8/9} \\
			& K=50                     & 0.923 & 0.868 & 0.879  &   & 0.960 & 0.918 & 0.885 & \checkmark & 0.921 & 0.853 & 0.810  & \checkmark & 2/3 &  \\
			& K=100                    & 0.875 & 0.838 & 0.778  & \checkmark & 0.930 & 0.900 & 0.859 & \checkmark & 0.873 & 0.829 & 0.778  & \checkmark & 3/3 &  \\ \midrule
			NBC                       & None                     & 0.450 & 0.362 & 0.263  & \checkmark & 0.496 & 0.422 & 0.323 & \checkmark & 0.497 & 0.341 & 0.267  & \checkmark & 3/3 & 3/3 \\ \midrule
			SNAC                      & None                     & 0.451 & 0.364 & 0.265  & \checkmark & 0.498 & 0.425 & 0.324 & \checkmark & 0.498 & 0.337 & 0.264  & \checkmark & 3/3 & 3/3 \\ \midrule
			NLC                       & \multirow{2}{*}{None}                     & 0.180 & 0.105 & 0.113  &  & 0.193 & 0.087 & 0.079 & \checkmark & 0.188 & 0.108 & 0.122  &   & 1/3 & \multirow{2}{*}{3/6} \\ 
			NLC+                      &                     & 0.660 & 0.588 & 0.547  & \checkmark & 0.672 & 0.643 & 0.644 &  & 0.479 & 0.481 & 0.442  & \checkmark & 2/3 &  \\ \midrule
			CC$_w$                        & None                     & 0.017 & 0.008 & 0.007  & \checkmark & 0.027 & 0.016 & 0.014 & \checkmark & 0.019 & 0.011 & 0.009   & \checkmark & 3/3 & 3/3 \\
			\bottomrule
		\end{tabular}
	}

\end{table*}

\begin{table*}[]
	\centering
	\caption{The average coverage values achieved by the sample sets ${\mathcal{S}}$, ${\mathcal{PSS_{+}}}$, and ${\mathcal{PSS}}$ on transformer-based language models for generation tasks. A checkmark indicates that the relationship $\mathcal{C}_{\mathcal{S}} > \mathcal{C}_{\mathcal{PSS_{+}}} > \mathcal{C}_{\mathcal{PSS}}$ is satisfied for the corresponding model. The final column shows the number of cases where this ground-truth relationship holds (out of two models).}
	\label{tab:rq2-gen}
	\resizebox{.80\textwidth}{!}{
		\begin{tabular}{ll|cccc cccc c c@{}}
			\toprule
			\multirow{2}{*}{\makecell[c]{\textbf{Coverage} \\ \textbf{Criteria}}} & {\multirow{2}{*}{\textbf{Config.}}} &
			\multicolumn{4}{c}{\textbf{Llama-2}} & \multicolumn{4}{c}{\textbf{T5}} & \multicolumn{2}{c}{\multirow{2}{*}{\makecell[c]{\textbf{Groundtruth} \\ \textbf{Satisfaction}}}} \\
			& & $\mathcal{S}$ & $\mathcal{PSS_{+}}$ & $\mathcal{PSS}$ &  & $\mathcal{S}$ & $\mathcal{PSS_{+}}$ & $\mathcal{PSS}$ &  &  &  \\ \midrule
			
			\multirow{3}{*}{NC} & T=0.25 & 0.902 & 0.890 & 0.890 &  & 0.935 & 0.932 & 0.927 & \checkmark & 1/2 & \multirow{3}{*}{5/6} \\
			& T=0.50 & 0.512 & 0.497 & 0.493 & \checkmark & 0.851 & 0.842 & 0.835 & \checkmark & 2/2 &  \\
			& T=0.75 & 0.281 & 0.243 & 0.226 & \checkmark & 0.602 & 0.589 & 0.571 & \checkmark & 2/2 &  \\ \midrule
			
			\multirow{3}{*}{KMNC} & K=10  & 0.925 & 0.884 & 0.861 & \checkmark & 0.913 & 0.859 & 0.828 & \checkmark & 2/2 & \multirow{3}{*}{6/6} \\
			& K=50  & 0.832 & 0.799 & 0.790 & \checkmark  & 0.802 & 0.742 & 0.702 & \checkmark & 2/2 &  \\
			& K=100 & 0.755 & 0.716 & 0.662 & \checkmark & 0.741 & 0.698 & 0.652 & \checkmark & 2/2 &  \\ \midrule
			
			NBC  & None & 0.362 & 0.285 & 0.201 & \checkmark & 0.404 & 0.276 & 0.199 & \checkmark & 2/2 & 2/2 \\ \midrule
			
			SNAC & None & 0.364 & 0.288 & 0.207 & \checkmark & 0.406 & 0.279 & 0.202 & \checkmark & 2/2 & 2/2 \\ \midrule
			
			NLC  & \multirow{2}{*}{None} & 0.112 & 0.066 & 0.071 &  & 0.121 & 0.071 & 0.083 &  & 0/2 & \multirow{2}{*}{1/4} \\
			NLC+ &  & 0.482 & 0.421 & 0.390 & \checkmark & 0.352 & 0.353 & 0.320 &  & 1/2 &  \\ \midrule
			
			CC$_w$   & None & 0.011 & 0.005 & 0.004 & \checkmark & 0.013 & 0.007 & 0.005 & \checkmark & 2/2 & 2/2 \\
			\bottomrule
		\end{tabular}
	}
\end{table*}

%% file: figure/figure_rq2.tex
\begin{figure}[t]
	\centering
	\includegraphics[scale=0.4]{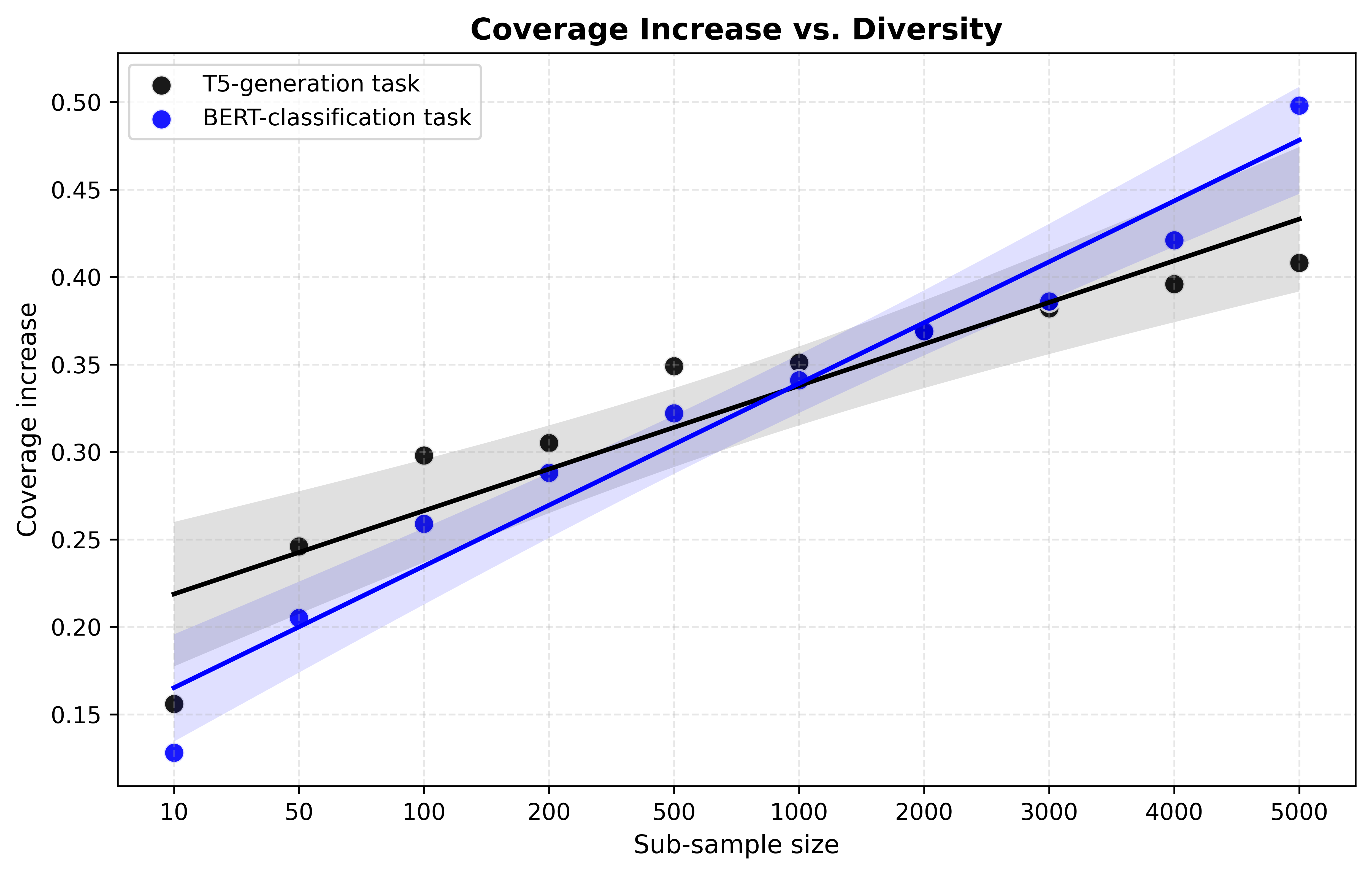}
	\vspace{-5pt}
	\caption{Coverage variation under fixed-size test suites generated from randomly perturbed sub-samples of different sizes.}
	\label{figure_rq2}
	\vspace{-15pt}
\end{figure}

%% file: chapter/evaluation3.tex
\subsection{Sample Fairness ( RQ3 )}

Sample fairness refers to the balance of categories within a test suite. It represents an important dimension of test-suite diversity and complements the distribution diversity emphasized in RQ2, as the two capture representativeness from the output label space and the input space, respectively. This research question, therefore, aims to explore whether neuron coverage criteria can effectively assess the sample fairness of a test suite. Specifically, given two test suites of the same size, we seek to determine whether coverage criteria can identify the test suite with better fairness, i.e., the one with a more balanced distribution of samples across categories.

\subsubsection{Experiment Design}  \vspace{0.5em}\noindent

\input{figure/figure3-3.tex}
\input{figure/figure3-32.tex}
To investigate RQ3, we design test suites with varying levels of sample fairness and analyze whether the coverage values produced by these test suites, when passed through the model, are positively correlated with their category diversity. Figure~\ref{figure3-32} provides an intuitive example of sample fairness, using a sentiment classification task as the context. In this example, circles and squares represent two different test suites, and different colors correspond to samples from different classes.
In the first test suite, samples are drawn from three classes with nearly equal representation, indicating balanced classes. This balance reflects good fairness or high category diversity. In contrast, the second test suite contains samples from only two classes, with a majority from the red class, a minority from the blue class, and none from the yellow class. This imbalance indicates poor fairness or low category diversity.

As illustrated in  Figure~\ref{figure3-33}, we construct varying levels of sample fairness.
We begin by defining a class-specific subset of size \(n\) for class \(c\) (denoted as $CSS$-$c$-$n$). From this, we can derive $CSS$-$c$-$2n$ by either combining $CSS$-$c$-$n$ with another subset of the same class or simply by doubling the size \(n\). Additionally, we define $CSS$-$dc$-$2n$ by combining $CSS$-$c$-$n$ with a subset from a different class, where ``$dc$'' stands for different classes.

\noindent \textit{Class-Specific Subset (denoted as $CSS$-$c$-$n$)}:

\[
CSS\text{-}c\text{-}n = \left\{ x_i \in \text{Dataset}_{\text{test}} \mid f(x_i) = y_i \text{ and } y_i = c \right\}; \quad \left| CSS\text{-}c\text{-}n \right| = n.
\]

This formula represents a subset of size \(n\) from the original sample \(S\), consisting exclusively of samples with the label \(c\). Here, \(f(x_i)\) represents the predicted label, and \(y_i\) is the ground truth label of the sample, with all \(y_i\) equal to \(c\). The subset $CSS$-$c$-$2n$ is created by either replacing \(n\) with \(2n\) in $CSS$-$c$-$n$ or by adding another $CSS$-$c$-$n$ subset to the original, effectively doubling the sample size. In this experiment, \(n\) is set to 2500.

\noindent \textit{Two-Class Subset (denoted as $CSS$-$dc$-$2n$) }:

\[
CSS\text{-}dc\text{-}2n = \left\{ x_i \in \text{Dataset}_{\text{test}} \mid f(x_i) = y_i \text{ and } y_i \in \{c_1, c_2\} \right\} ; \quad \left| \{y_i = c_1\} \right| = \left| \{y_i = c_2\} \right| = n.
\]

The $CSS$-$dc$-$2n$ subset is constructed by combining two class-specific subsets, each of size \(n\), from classes \(c_1\) and \(c_2\). Here, \(f(x_i)\) represents the predicted label, and \(y_i\) is the ground truth label, where \(y_i\) can be either \(c_1\) or \(c_2\). The total size of $CSS$-$dc$-$2n$ is \(2n\), with exactly \(n\) samples from each class. As above, \(n\) is set to 2500.

After constructing the three test suites, we rank their fairness as \( CSS\text{-}dc\text{-}2n  > CSS\text{-}c\text{-}2n  > CSS\text{-}c\text{-}n \). This is because CSS-c-n contains 2,500 samples from a single class, while CSS-c-2n doubles the sample size but still draws from the same class, offering only a slight diversity gain. By contrast, CSS-dc-2n introduces an equal number of samples from a different class, resulting in the highest fairness. We then apply these test suites to transformer-based language models and compute coverage values under each criterion. Using the CC method~\cite{ji2023cc} as a reference, the ground-truth ranking of coverage values is established as \( \mathcal{C}_{\text{dc-2n}} > \mathcal{C}_{\text{c-2n}} > \mathcal{C}_{\text{c-n}} \). Finally, we evaluate how closely each neuron coverage criterion aligns with this ground truth.

\subsubsection{Results} \vspace{0.5em}\noindent

\input{table/table_rq3}

Table~\ref{tab:rq3} presents the average coverage values achieved by the sample sets \texttt{CSS-c-n}, \texttt{CSS-c-2n}, and \texttt{CSS-dc-2n} across various transformer-based language models. The main columns represent the models, further subdivided into three sub-columns indicating the respective sample sets. The table is designed to evaluate whether the coverage values adhere to the expected relationship $\mathcal{C}_{\text{c-n}} < \mathcal{C}_{\text{c-2n}} < \mathcal{C}_{\text{dc-2n}}$, with a checkmark indicating when this condition is satisfied. The final column summarizes the proportion of cases where the expected relationship is met.
For instance, in the first row under the NC criterion with a threshold of $T=0.25$, the coverage values of $0.958$, $0.962$, and $0.966$ for the LLama-2 model correspond to the sample sets \texttt{CSS-c-n}, \texttt{CSS-c-2n}, and \texttt{CSS-dc-2n}, respectively. Since $0.958 < 0.962 < 0.966$ satisfies the expected relationship, a checkmark is placed accordingly. However, for the BERT model, the values $0.999$, $1.000$, and $1.000$ do not satisfy the relationship since the latter two values are equal; hence, no checkmark is recorded.

When examining the NC criterion at thresholds $T=0.25$, $T=0.50$, and $T=0.75$, the relationship $\mathcal{C}_{\text{c-n}} < \mathcal{C}_{\text{c-2n}} < \mathcal{C}_{\text{dc-2n}}$ is generally maintained, particularly at higher thresholds, with $7$ out of $9$ configurations conforming to the expected pattern. All three models adhere to this relationship at $T=0.75$. Even in the two cases where the relationship is not fully satisfied, the first inequality, $\mathcal{C}_{\text{c-n}} < \mathcal{C}_{\text{c-2n}}$, is still maintained (e.g., $0.999 < 1.000$ and $0.997 < 0.999$). This suggests that at lower thresholds, the shift from $\mathcal{C}_{\text{c-n}}$ to $\mathcal{C}_{\text{c-2n}}$ is consistently observed, while the more subtle shift from $\mathcal{C}_{\text{c-2n}}$ to $\mathcal{C}_{\text{dc-2n}}$ is harder to discern.
For the KMNC, NBC, SNAC, and CC criteria, there is a strong alignment with the expected relationship, consistently satisfying the expected pattern across all models and configurations, achieving an overall success rate of $18/18$. This consistency underscores the reliability of these criteria in evaluating the fairness of the sample sets.
The NLC family also achieves a high rate of adherence to the ground truth, with $5$ out of $6$ configurations meeting the expected relationship. Specifically, for NLC, all three configurations ($3/3$) are satisfied. This is likely because NLC is based on variance, which might decrease when adding certain samples that are very close in the input space, thereby lowering the NLC coverage value. However, during the fairness evaluation process, where new samples—often from different categories—are added, the close proximity effect is minimal, resulting in consistent outcomes. For NLC+, $2$ out of $3$ configurations meet the expected relationship. In the unsatisfied example ($0.557$, $0.551$, $0.695$), we observe that $0.557 > 0.551$. As previously noted, NLC+ is sensitive to the order of sample inputs, leading to a scenario where the first sample set, due to input order, may result in a higher coverage value. However, for $\mathcal{C}_{\text{c-2n}}$ and $\mathcal{C}_{\text{dc-2n}}$, where the latter involves the addition of entirely new categories, the impact of order-induced errors becomes negligible.

In summary, the table shows that the relationship $\mathcal{C}_{\text{c-n}} < \mathcal{C}_{\text{c-2n}} < \mathcal{C}_{\text{dc-2n}}$ holds in 30 out of 33 cases. In the remaining three cases, a partial relation is satisfied, either $\mathcal{C}_{\text{c-n}} < \mathcal{C}_{\text{c-2n}}$ or $\mathcal{C}_{\text{c-n}} < \mathcal{C}_{\text{dc-2n}}$. The NBC, SNAC, KMNC, and CC criteria show the most consistent adherence to this pattern, while NC and NLC mostly conform. In cases where NC does not satisfy the relationship, it is generally due to the low threshold, which allows $\mathcal{C}_{\text{c-2n}}$ to reach near-complete coverage values (close to $1$). As a result, even when $\mathcal{C}_{\text{dc-2n}}$ introduces new categories, it fails to uncover new, previously uncovered information. Overall, NC typically satisfies $\mathcal{C}_{\text{c-n}} < \mathcal{C}_{\text{c-2n}} \leq \mathcal{C}_{\text{dc-2n}}$. For cases where NLC does not meet the expected relationship, this is attributed to the fact that datasets from the same category, even with different sample sizes, exhibit minimal semantic differences. Consequently, smaller sample sets, depending on their input order, might yield higher coverage values. Nonetheless, the NLC family generally satisfies $\mathcal{C}_{\text{c-n}} < \mathcal{C}_{\text{dc-2n}}$ and $\mathcal{C}_{\text{c-2n}} < \mathcal{C}_{\text{dc-2n}}$, as the addition of new categories effectively counterbalances potential order-induced errors.

\begin{tcolorbox}[title=Finding 3:,colbacktitle=gray!50, coltitle=black,fonttitle=\bfseries,boxrule=0.8pt]
	{\textbf{Neuron coverage criteria effectively assess the sample fairness of a test suite in 30 out of 33 cases. The remaining three instances still satisfy half of the expected relation.}}
\end{tcolorbox}

\begin{tcolorbox}[title=Implication 3:,colbacktitle=gray!50, coltitle=black,fonttitle=\bfseries,boxrule=0.8pt]
	{ 
		\begin{compactitem}[$\bullet$]
                \item  \textbf{Usage and Design.} Experimental results indicate that existing coverage criteria already provide reasonable effectiveness in assessing the category diversity of test suites. This suggests that category diversity may be less pressing as a direction for new criterion design, though further exploration into how multiple criteria can be strategically integrated may still offer additional benefits.
		\end{compactitem}
	}
\end{tcolorbox}

%% file: figure/figure3-3.tex
\begin{figure}[t]
	\centering
	\includegraphics[scale=0.18]{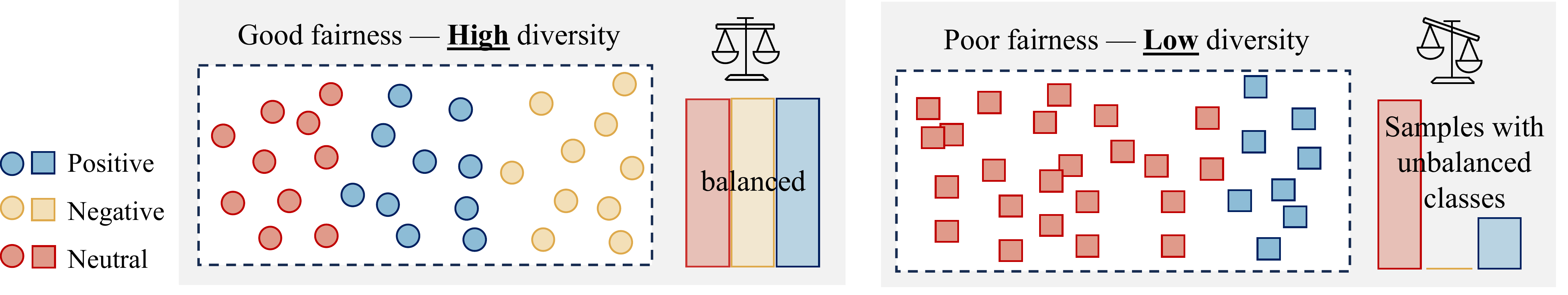}
	 \vspace{-5pt}
	\caption{An illustration of sample fairness in sentiment classification. This example compares test suite of equal size. The first sub-figure demonstrates good fairness, with a balanced distribution across all categories, while the second sub-figure illustrates poor fairness, characterized by missing and imbalanced classes.}
	\label{figure3-32}
 \vspace{-8pt}
\end{figure}

%% file: figure/figure3-32.tex
\begin{figure}[t]
	\centering
	\includegraphics[scale=0.17]{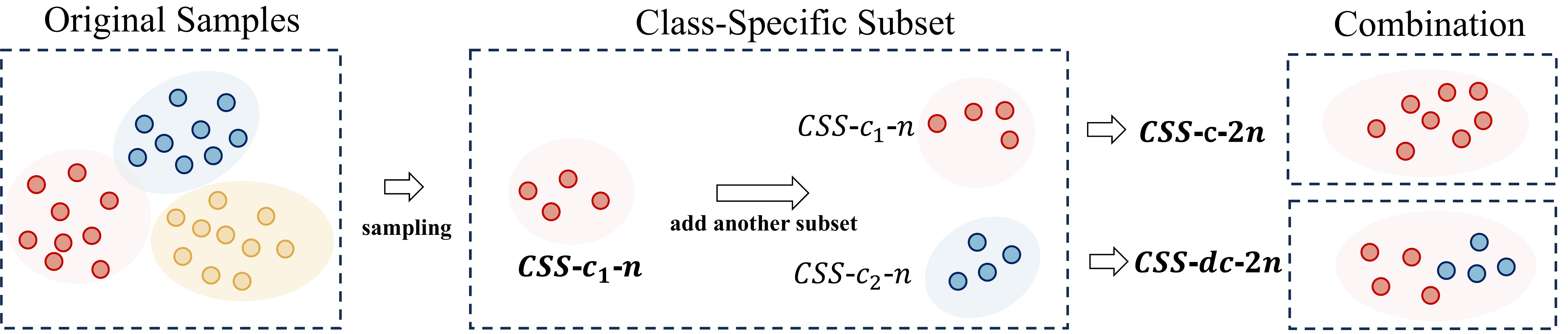}
	 \vspace{-5pt}
	\caption{Process for constructing the $CSS$-$c$-$2n$ and $CSS$-$dc$-$2n$ test suites: Randomly select two subsets of class $c_1$ and one subset of class $c_2$. $CSS$-$c$-$2n$ is formed by combining the $c_1$ subsets ($c_1 + c_1$), while $CSS$-$dc$-$2n$ combines $c_1$ and $c_2$ ($c_1 + c_2$).}
	\label{figure3-33}
  \vspace{-10pt}
\end{figure}

%% file: table/table_rq3.tex
\begin{table*}[]
	\centering
	\caption{The average coverage values achieved by the sample sets $CSS$-$c$-$n$, $CSS$-$c$-$2n$, and $CSS$-$dc$-$2n$ on transformer-based language models. A checkmark indicates that the relationship \(  \mathcal{C}_{\text{c-n}} < \mathcal{C}_{\text{c-2n}} < \mathcal{C}_{\text{dc-2n}} \) is satisfied. The final column shows the number of cases where this ground truth relationship holds.}
	\label{tab:rq3}
	\resizebox{.95\textwidth}{!}{
		\begin{tabular}{ll|cccc cccc cccc c c@{}}
			\toprule
			\multirow{2}{*}{\makecell[c]{\textbf{Coverage} \\ \textbf{Criteria}}} &  {\multirow{2}{*}{\textbf{Config.}}} &
			\multicolumn{4}{c}{\textbf{LLama-2}} & \multicolumn{4}{c}{\textbf{BERT}} & \multicolumn{4}{c}{\textbf{T5}} & \multicolumn{2}{c}{\multirow{2}{*}{\makecell[c]{\textbf{Groundtruth} \\  \textbf{Satisfaction}}}} \\ 
			&   & $c$-$n$  & $c$-$2n$ & $dc$-$2n$ &   & $c$-$n$  & $c$-$2n$ & $dc$-$2n$ &   & $c$-$n$  & $c$-$2n$ & $dc$-$2n$ &   &  &  \\ \midrule
			\multirow{3}{*}{NC}       & T=0.25                   & 0.958 & 0.962 & 0.966  & \checkmark & 0.999 & 1.000 & 1.000 &  & 0.973 & 0.986 & 0.991  & \checkmark & 2/3 & \multirow{3}{*}{7/9} \\
			& T=0.50                   & 0.581 & 0.593 & 0.603  & \checkmark & 0.997 & 0.999 & 0.999 &  & 0.951 & 0.955 & 0.970  & \checkmark & 3/3 &  \\
			& T=0.75                   & 0.339 & 0.347 & 0.371  & \checkmark & 0.590 & 0.675 & 0.830 & \checkmark & 0.671 & 0.673 & 0.701  & \checkmark & 3/3 &  \\ \midrule
			\multirow{3}{*}{KMNC}     & K=10                     & 0.952 & 0.964 & 0.989  & \checkmark & 0.944 & 0.961 & 0.996 & \checkmark & 0.933 & 0.945 & 0.987  & \checkmark & 3/3 & \multirow{3}{*}{9/9} \\
			& K=50                     & 0.819 & 0.876 & 0.923  & \checkmark & 0.751 & 0.887 & 0.960 & \checkmark & 0.797 & 0.847 & 0.923  & \checkmark & 2/3 &  \\
			& K=100                    & 0.801 & 0.823 & 0.875  & \checkmark & 0.775 & 0.840 & 0.930 & \checkmark & 0.716 & 0.794 & 0.873  & \checkmark & 3/3 &  \\ \midrule
			NBC                       & None                     & 0.313 & 0.327 & 0.450  & \checkmark & 0.399 & 0.420 & 0.496 & \checkmark & 0.308 & 0.350 & 0.491  & \checkmark & 3/3 & 3/3 \\ \midrule
			SNAC                      & None                     & 0.291 & 0.327 & 0.451  & \checkmark & 0.411 & 0.424 & 0.498 & \checkmark & 0.329 & 0.366 & 0.498  & \checkmark & 3/3 & 3/3 \\ \midrule
			NLC                       & None                     & 0.093 & 0.106 & 0.180  & \checkmark & 0.112 & 0.118 & 0.193 & \checkmark & 0.107 & 0.109 & 0.188  & \checkmark & 3/3 & \multirow{2}{*}{5/6} \\  
			NLC+                      & None                     & 0.557 & 0.551 & 0.695  &  & 0.531 & 0.602 & 0.668 & \checkmark & 0.394 & 0.401 & 0.479  & \checkmark & 2/3 &  \\ \midrule
			CC$_w$                        & None                     & 0.003 & 0.004 & 0.017  & \checkmark & 0.009 & 0.11 & 0.027 & \checkmark & 0.012 & 0.014 & 0.019  & \checkmark & 3/3 & 3/3 \\
			\bottomrule
		\end{tabular}
	}
	\vspace{-10pt}
\end{table*}

%% file: chapter/evaluation4.tex
\subsection{Input Generation Guidance ( RQ4 )}
Input generation refers to the process in software testing where new test inputs are generated from an initial sample set using various mutation techniques. This research question aims to explore whether guiding input generation with the goal of increasing neuron coverage criteria can produce more valuable test cases. Specifically, it examines whether incorporating coverage criteria into conventional mutation methods can lead to the generation of a higher number of fault-revealing samples.
\subsubsection{Experiment Design}  \vspace{0.5em}\noindent

To address RQ4, we conduct fuzz testing on a text-based task guided by neuron coverage criteria. Coverage-guided fuzzing approaches typically use coverage as the fitness function, supplemented by seed scheduling and mutation strategies. Our objective is to determine whether coverage-guided fuzzing can generate more adversarial examples—inputs that are semantically similar to the original samples but yield different outputs. While numerous fuzz testing methods exist, most focus on image tasks. Since our work targets text datasets, we first identify common perturbation strategies for text. Simple word swaps are unsuitable for this purpose, as they can significantly alter the semantics of sentences. For example, transforming \emph{I saw a man with a telescope} into \emph{I saw a telescope with a man} drastically changes the meaning. Thus, word swaps are not appropriate for introducing noise into text data. Instead, we employ techniques such as synonym replacement, tense modification, and repetition as mutation operators. By applying these mutation operators, we perform random mutations on the initial dataset to generate new sample sets, establishing a baseline for sample generation without coverage guidance.

\input{algorithm/alg1.tex}

With the basic sample generation algorithm established, we develop a coverage-guided input generation algorithm to strategically select mutated samples based on coverage criteria. The detailed method is presented in Algorithm ~\ref{alg1}. In Lines 1-2, the algorithm begins by initializing the input parameters: the initial seed set \(S\), the target neural network \(DNN\), the coverage criterion \(C\), and a custom parameter \(N\) that limits the number of generated samples. In this experiment, \(S\) refers to the original samples, where a total of 1,000 inputs are randomly selected from the test dataset to serve as fuzzing seeds, and \(N\) is set to 10,000. The output set \(F\) is initialized as empty, and the seed set \(S\) is loaded into the queue \(Q\). In Lines 3-5, the algorithm enters a loop where it selects the next seed \(s\) from the queue \(Q\) (line 5) and applies the \textit{TextMutate} function to generate a set \(T\) of mutated samples (line 6). In Line 7, the mutated samples \(T\) are evaluated against the \(DNN\) using the coverage criterion \(C\). The \textit{Run} function returns the coverage information \(cov\), which guides the selection of further mutations. In Lines 8-16, each sample \(s'\) in \(T\) is processed as follows: the sample counter \(i\) is incremented (line 8) and checked against the limit \(N\) (line 9). If \(i\) exceeds \(N\), the process stops. If a sample is adversarial (\textit{isAdversarial}), it is added to the output set \(F\) (lines 10-11). If a sample increases coverage (\textit{isCoverageGain}), it is added to the queue \(Q\) for further mutation (lines 12-13). In Line 17, the loop continues until the sample generation limit \(N\) is reached or no more seeds are available for mutation, at which point the algorithm terminates. This coverage-guided approach ensures that the algorithm focuses on samples that either reveal new behaviors or improve coverage, optimizing the efficiency of the fuzz testing process. 
Given the large size of transformer-based language models, we select neuron-level coverage criteria with the lowest computational complexity, namely NC, NBC, and SNAC, as guiding indicators. During the generation process, if any of these coverage criteria increases, the function \texttt{isCoverageGain(cov)} returns true.
Overall, we adopt the random mutation strategy with the same operators as the baseline. In addition, we evaluate the effect of applying three coverage criteria individually, as well as their combination. We then compare the results across these four methods by measuring the proportion of adversarial samples generated from the initial sample set for classification tasks, and by analyzing the BLEU drop for generation tasks, both based on 10,000 newly generated samples.

\input{table/table_rq4.tex}

\subsubsection{Results} \vspace{0.5em}\noindent

Table~\ref{tab:rq4} and Table~\ref{tab:rq4-gen} summarize the results under different mutation strategies, reporting the proportion of adversarial samples for classification tasks and the BLEU drop for generation tasks, respectively. As shown in Table~\ref{tab:rq4}, the baseline random mutation yields 1,589 adversarial samples out of 10,000 inputs (15.89\%). Incorporating coverage criteria individually leads to modest but noticeable improvements: With $A$ (NC) generates 1,604 samples (16.04\%), With $B$ (NBC) 1,815 samples (18.15\%), and With $C$ (SNAC) 1,801 samples (18.01\%). Among the three, $B$ and $C$ show slightly stronger effects than $A$, suggesting that different criteria capture complementary aspects of neuron activations. The combined strategy With $A{+}B{+}C$ achieves the highest proportion, producing 2,045 adversarial samples (20.45\%), outperforming random mutation by 4.56 percentage points and consistently surpassing the individual criteria. 
A similar trend is observed in Table~\ref{tab:rq4-gen}. While random mutation reduces BLEU by 7.11\%, coverage-guided strategies achieve larger drops, with the combined variant ($A{+}B{+}C$) reaching the maximum decrease of 10.65\%. The stronger BLEU degradation indicates that coverage-guided mutations introduce more semantically disruptive perturbations, thereby uncovering failure cases that random mutation is less likely to expose.

In addition to effectiveness, Tables~\ref{tab:rq4} and \ref{tab:rq4-gen} report the empirical wall-clock time of coverage-guided mutation.
For classification tasks (Table~\ref{tab:rq4}), incorporating a single coverage criterion introduces a moderate computational overhead compared to random mutation. Specifically, NC-, NBC-, and SNAC-guided strategies require approximately 97 s, 115 s, and 157 s, respectively, while the combined strategy ($A{+}B{+}C$) incurs the highest cost at around 265,s. This increase is expected, as integrating more coverage criteria entails additional coverage computation and selection steps during mutation.
For generation tasks (Table~\ref{tab:rq4-gen}), the time cost is substantially higher, ranging from approximately 2,146 s for NC-guided mutation to about 7,702 s for the combined strategy. This roughly one-order-of-magnitude increase compared to classification tasks is mainly attributable to the substantially larger model sizes and the higher cost of computing coverage over longer output sequences in generation models.
Despite this overhead, the observed wall-clock times remain practical under the experimental setting described in Section~\ref{settings}, where all experiments are conducted with a batch size of 32 on the same hardware configuration. It is also worth noting that these time costs are not inherent limitations of coverage-guided testing. In practice, they can be further reduced through engineering optimizations such as increasing batch size, parallelizing coverage computation, or leveraging more powerful hardware accelerators. Therefore, the reported results serve as a realistic empirical reference rather than an upper bound on the efficiency of coverage-guided fuzzing.

Overall, these results demonstrate that coverage guidance consistently expands the adversarial search space beyond unguided random perturbations, with stable improvements observed across both classification and generation tasks. While the improvement brought by any single coverage criterion is moderate, this is consistent with our earlier findings that coverage signals tend to be weak but directionally correct. When leveraged within an iterative fuzzing process, such signals are progressively amplified through accumulation and pruning: perturbations that fail to induce new internal behaviors are filtered out, while coverage-increasing mutations are retained and reused, effectively narrowing the search space toward error-prone regions.
Moreover, combining multiple criteria yields the most robust gains, as NC, NBC, and SNAC capture complementary aspects of model behavior. Their synergy amplifies weak but consistent signals and provides stronger guidance than any individual criterion alone. As a result, coverage-guided mutation not only increases the proportion of adversarial samples in classification tasks, but also more effectively degrades generation quality, reinforcing its advantage over blind random mutation and highlighting its practical value for improving the efficiency of neural network testing.

\begin{tcolorbox}[title=Finding 4:,colbacktitle=gray!50, coltitle=black,fonttitle=\bfseries,boxrule=0.8pt]
	{ \textbf{Under the guidance of neuron-level coverage criteria, the sample generation is slightly enhanced.
    Specifically, the improvement is as low as 0.15\% for individual criterion, and only by combining three criteria does it produce 4.56\% more adversarial samples compared to basic methods in a set of 10,000 samples.
	} }
\end{tcolorbox}

\begin{tcolorbox}[title=Implication 4:,colbacktitle=gray!50, coltitle=black,fonttitle=\bfseries,boxrule=0.8pt]
	{ 
		\begin{compactitem}[$\bullet$]
			\item \textbf{Usage.} In practice, combining diverse coverage criteria can, to a certain extent, more effectively constrain the search space of random mutations to guide the discovery of error-revealing inputs. However, due to cost-effectiveness concerns, practical applications must carefully weigh the computational overhead against the potential benefits.		
		\end{compactitem}
	}
\end{tcolorbox}

%% file: algorithm/alg1.tex
\begin{algorithm}[t]
	\caption{Coverage-Guided Input Generation}
	\begin{algorithmic}[1]
		\State \textbf{Input}: \textit{S}: Initial Seed Set, \textit{DNN}: Target Neural Network, \textit{C}: Coverage Criterion, \textit{N}: Maximum Number of Generated Samples
		\State \textbf{Output}: Set of adversarial examples $F \gets \emptyset$
		\State $Q \gets S$ \Comment{Initialize the queue with the seed set}
		\State $i \gets 0$ \Comment{Initialize the sample counter}
		\While{$s \gets \text{\textbf{SelectNext}}(Q)$} \Comment{Select the next seed from the queue}
		\State $T \gets \text{\textbf{TextMutate}}(s)$ \Comment{Generate mutated text samples from seed $s$}
		\State $cov \gets \text{\textbf{Run}}(\textit{DNN}, T, C)$ \Comment{Evaluate coverage of mutated samples}
		\ForAll{$s' \in T$} \Comment{Iterate through each mutated sample}
		\State $i \gets i + 1$ \Comment{Increment the sample counter}
		\If{$i > N$} \Comment{Stop if the number of generated samples exceeds the limit}
		\State \textbf{break}
		\ElsIf{$\text{\textbf{isAdversarial}}(s')$} \Comment{Check if the sample is adversarial}
		\State $F \gets F \cup \{s'\}$ \Comment{Add adversarial sample to the output set}
		\ElsIf{$\text{\textbf{isCoverageGain}}(cov)$} \Comment{Check if the sample increases coverage}
		\State $Q \gets Q.\text{\textbf{append}}(s')$ \Comment{Add the sample to the queue for further exploration}
		\EndIf
		\EndFor
		\EndWhile
	\end{algorithmic}
	\label{alg1}
\end{algorithm}

%% file: table/table_rq4.tex
% Please add the following required packages to your document preamble:
% \usepackage{graphicx}

\begin{table*}[t]
	\centering

	\caption{Proportion of adversarial samples on classification tasks: comparison of random mutation and its variants with different coverage-guided strategies (A, B, C), where
		A, B, and C correspond to NC, NBC, and SNAC, respectively. 
		Wall-clock time is reported in seconds (s).}
	\label{tab:rq4}
	\resizebox{0.95\textwidth}{!}{
		\begin{tabular}{c|ccccc}
			\toprule
			\textbf{Input Generation Strategy} 
			& Random Mutation 
			& With $A$ 
			& With $B$ 
			& With $C$ 
			& With $A{+}B{+}C$ \\ 
			\midrule
			\textbf{Adversarial Proportion} 
			& 1589 / 10000 
			& 1604 / 10000 
			& 1815 / 10000 
			& 1801 / 10000 
			& \textbf{2045 / 10000} \\ 
			\textbf{Wall-clock Time (s)} 
			& -- 
			& $\sim$97.24 
			& $\sim$114.71 
			& $\sim$157.27 
			& $\sim$265.44 \\
			\bottomrule
		\end{tabular}
	}
	\vspace{-8pt}
\end{table*}

\begin{table*}[t]
	\centering

	\caption{$\Delta$BLEU (↓) on generation tasks: comparison of random mutation and its variants with different coverage-guided strategies (A, B, C), where A, B, and C correspond to NC, NBC, and SNAC, respectively. Wall-clock time is reported in seconds.}
	\label{tab:rq4-gen}
	\resizebox{.90\textwidth}{!}{
		\begin{tabular}{c|ccccc}
			\toprule
			\textbf{Input Generation} 
			& Random Mutation 
			& With $A$ 
			& With $B$ 
			& With $C$ 
			& With $A{+}B{+}C$ \\
			\midrule
			\textbf{$\Delta$BLEU} 
			& -7.11\% 
			& -7.56\% 
			& -7.94\% 
			& -7.61\% 
			& \textbf{-10.65\%} \\
			
			\textbf{Wall-clock Time (s)} 
			& -- 
			& $\sim$2145.63 
			& $\sim$2990.14 
			& $\sim$3841.55 
			& $\sim$7701.95 \\
			\bottomrule
		\end{tabular}
	}
	\vspace{-8pt}
\end{table*}

%% file: chapter/relatedwork.tex
\section{Related work}
\label{sec:related_work}
\subsection{Transformer-based Model}

The Transformer architecture is one of the most popular neural network structures and core technology behind the widely acclaimed ChatGPT\cite{OpenAI2023}. 
The realm of deep learning underwent a revolutionary shift with the introduction of the Transformer architecture by Vaswani et al \cite{vaswani2017attention}. Originally designed to address challenges in sequence transduction tasks such as machine translation, the Transformer model leverages self-attention mechanisms allowing it to weigh the significance of different parts of an input sequence relative to each other.
Following this landmark work, numerous Transformer-based models have been proposed, setting new benchmarks across various tasks. BERT\cite{devlin2018bert} stands out as a prominent example, introducing the concept of bidirectional training of Transformers for language understanding. BERT's pre-training on vast corpora and fine-tuning on specific task paradigms has become a de facto standard in numerous NLP tasks.
Expanding upon BERT, models like RoBERTa\cite{liu2019roberta} modify aspects of BERT's training regimen to achieve even better performance. GPT\cite{radford2018improving} and its subsequent versions\cite{radford2019language,brown2020language,ouyang2022training,bubeck2023sparks} showcase the prowess of large-scale, generative pre-trained Transformers, particularly in language generation tasks.
Attention mechanisms, a key component of Transformer models, have undergone refinements. XLNet\cite{yang2019xlnet}, for instance, integrates permutation-based training to capture bidirectional context more effectively than traditional autoregressive models. Meanwhile, efforts like T5\cite{raffel2020exploring} revisit the paradigm by treating every NLP challenge as a text-to-text problem, showcasing the flexibility and robustness of Transformer models.
Apart from the advances in core model architectures, there have also been significant strides in model efficiency. Efforts towards model efficiency have resulted in variants like ALBERT\cite{lan2019albert}, DistilBERT\cite{sanh2019distilbert}, and TinyBERT\cite{jiao2019tinybert}, offering pruned or distilled versions of larger Transformers. These models ensure that the benefits of Transformer architectures can be leveraged even in resource-constrained environments, broadening their applicability and impact in real-world scenarios.

In summary, Transformer-based models have significantly reshaped the landscape of NLP and continue to lead the way in state-of-the-art performance across a myriad of tasks. However, due to the complexity of these models, there is still a lack of comprehensive testing methods available to evaluate them.
% Language models are advanced artificial intelligence models designed to process and generate human-like text. Large language models (LLMs) ,which typically refer to large-sized pre-trained language models, have rapidly developed in recent years. The emergence of GPT-2 ushered in the era of big language models. Compared with early pre-trained language models (PLMs) such as Bert, GPT and GPT-2, gpt-3 exhibits astonishing ability in solving a series of complex tasks. With the advent of Chatgpt\cite{OpenAI2023} and gpt-4, the field of artificial intelligence and natural language processing is undergoing revolutionary changes.

% Recently, PLMs and LLMs have been widely discussed and surveyed. However, most of the work focuses on their applications with little testing related work. We apply the neuron coverage criteria to language models and analyze its effectiveness. Through testing, we can verify the robustness of different methods and analyze the performance of the model.

\subsection{Neuron Coverage Testing}

As deep learning models, particularly deep neural networks (DNNs), have gained widespread acceptance in diverse domains, ranging from image recognition to natural language processing, the imperative to ensure the reliability of these models has become increasingly crucial.  
Unlike traditional software, which has well-established testing criteria\cite{ammann2016introduction}, the unique characteristics of DNNs necessitate the formulation of new testing standards. Neuron coverage emerges as one such criterion designed to gauge the thoroughness of DNN testing, with DeepXplore\cite{deepxplore} pioneering as the first white-box testing framework for real-world DNN systems, introducing the concept of neuron coverage for the first time.
Considering that the granularity of the neuron coverage proposed in DeepXplore is relatively coarse, DeepCover\cite{deepcover} and DeepGauge\cite{deepgauge} are proposed accordingly and demonstrate better performance. Inspired by traditional combination testing methods and path-oriented testing methods, DeepCT\cite{deepct} and DeepPath\cite{deeppath} were subsequently proposed, achieving higher coverage in terms of testing coverage standards. In order to accomplish the task of automatically identifying the confusion and deviation of DNN-based image classifiers that cannot be realized by existing DNN testing technology, the White-box testing framework DeepInspect\cite{deepinspect} was proposed. The CNN-based test coverage criteria above do not have good portability for RNN. Therefore, the automation test framework DeepCruiser\cite{deepcruiser} which is based on the RNN test coverage criteria, is proposed and can effectively detect defects in the RNN-based system. Furthermore, to study a single input, Kim et al. propose a fine-grained testing adequacy framework SADL\cite{surpriseadequacy} and introduce two indicators: surprise adequacy (SA) and surprise coverage (SC), which play an important role in distinguishing adversarial samples and guiding model retraining.

However, most of the existing coverage criteria are used for DNN and image-related tasks, with a few used for NLP tasks but have only been tested on RNN. Accordingly, we apply neuron coverage testing methods to transformer-based language models to evaluate their effectiveness and the robustness of the model.

%% file: chapter/discussion.tex
\section{discussion}
\label{sec:discussion}

\noindent \textbf{Implications.}
Our findings suggest several practical implications for the use and future design of neuron coverage criteria in Transformer-based models. First, for error detection (RQ1), the marginal coverage gains from adversarial samples over benign noise highlight the limited practical value of existing criteria; future designs should explicitly distinguish erroneous from normal samples by amplifying error-related semantics while suppressing noise-like perturbations. Second, regarding distributional diversity (RQ2), most existing criteria—except the order-sensitive NLC series—can already be applied, but those with thresholds require careful tuning; future criteria should remain robust to input order while preserving sensitivity to diversity. Third, for category diversity (RQ3), current criteria show adequate performance in this dimension, implying that future design efforts might consider focusing more on other aspects while still keeping this one in view. Finally, for input generation (RQ4), combining multiple coverage criteria provides a more effective guide than individual ones, shrinking the search space of random mutations and enabling more efficient discovery of error-revealing inputs. 
To further illustrate the implications, we summarize two potential ways in which coverage criteria could be applied to improve the reliability of Transformer-based NLP systems: coverage-guided sample selection and coverage-guided fuzzing for input generation.  

\begin{compactitem}[$\bullet$]
    \item \textbf{Coverage-guided Sample Selection.} 
    Coverage information may assist in selecting representative samples under resource constraints. 
    For example, in data annotation for classification tasks, when only a limited number of samples can be labeled, coverage could help identify subsets with more balanced category distribution (as suggested by RQ3), thereby improving the cost-effectiveness of annotation. 
    Similarly, in fine-tuning for text generation tasks, when large training corpora cannot be fully utilized due to computational limits, coverage could guide the selection of samples with more balanced distribution (as suggested by RQ2), reducing data skewness and enhancing fine-tuning efficiency.  

    \item  \textbf{ Coverage-guided Fuzzing for Input Generation.} 
    Coverage may also serve as a feedback signal in fuzzing-based input generation. 
    Rather than applying random perturbations, coverage-guided fuzzing could prioritize mutations that explore previously under-tested regions of the model(as suggested by RQ4). 
    In text generation tasks, for instance, such perturbations may be more likely to trigger factual inconsistencies or other error-revealing behaviors, thereby producing challenging test cases. 
    This targeted augmentation has the potential to support more effective error detection and robustness evaluation compared to unguided approaches. 

\end{compactitem}
 
\noindent \textbf{Model Scale and Coverage Effectiveness.} Although the complexity of most coverage criteria is polynomial, the computational cost is positively correlated with network depth and the number of neurons per layer, leading to varying degrees of growth as models expand. Therefore, analyzing the sensitivity of coverage criteria across models of varying sizes is essential. Based on our experiments covering models with vastly different parameter sizes, including the Encoder-only BERT (110M), the Encoder-Decoder T5 (220M), and the Decoder-only LLaMA (7B), we demonstrate that criterion efficacy remains consistent across model scales. Specifically, in terms of Error-Revealing Capability (RQ1), all models exhibited nearly identical trends. For Distribution Diversity (RQ2) and Sample Fairness (RQ3), the criteria effectively reflected test suite properties with comparable success rates across sizes (RQ2: 8/11, 9/11, 9/11; RQ3: 9/11, 11/11, 10/11 for BERT, T5, and LLaMA, respectively). The slightly lower success rate observed in the smallest model (BERT) is attributable to coverage saturation,where coverage approaches 100\% even on low-diversity suites, rather than a lack of sensitivity. These findings confirm that, within our experimental range (110M to 7B), the validity of coverage criteria is not compromised by the exponential growth of parameters.

However, to ensure reliability, our current approach employs average pooling over all internal representations, considering the model's entire information flow. While effective, this imposes a substantial computational overhead that grows significantly with model size. Consequently, for future testing of ultra-large models, research should pivot towards identifying  representative subsets within the internal representations as testing targets. This strategy aims to maintain the efficacy of coverage criteria while significantly enhancing computational efficiency.

\section{Threats to Validity}

\subsection{External Validity}
External validity threats might arise from the scope of our experiments, which are conducted on six language models and six datasets. 
To mitigate this risk, we include the three most widely used transformer model architectures: encoder-only, decoder-only, and encoder-decoder. 
These choices make our experiments broadly representative of transformer-based language models. 
Nevertheless, in future research, we will explore whether our findings hold across a wider range of model architectures, including hybrid networks that integrate transformers with convolutional networks.

\subsection{Internal Validity}
Internal validity threats could emerge from the randomness in model fine-tuning. 
To mitigate this threat, we employed five different seeds during the fine-tuning process and averaged the results to ensure that the randomness had minimal impact on our conclusions.
\noindent  Another potential threat arises from the data preprocessing strategy, specifically the truncation of input texts. While imposing maximum text length constraints is a standard practice to optimize model inference speed across varying language tasks, it inherently poses a risk of losing semantic information located at the end of sequences. To mitigate this risk, we struck a balance between efficiency and information integrity. As detailed in Table 3, for the majority of datasets, the maximum length was set to 128, which significantly exceeds their average lengths (ranging from 20.3 to 39.8), thereby affecting only a negligible portion of long-tail samples. For the dataset characterized by longer sequences (i.e., AGnews), we extended the limit to 256. Although this is closer to its average length of 233, preliminary experiments on the test set demonstrated that this length is sufficient for the model to capture critical semantic features. The performance discrepancy (e.g., accuracy) between inputs truncated at 256 and full-length inputs was observed to be less than 2\%. Therefore, we believe our truncation strategy minimizes semantic loss while ensuring experimental feasibility.

%% file: chapter/conlusion.tex
\section{Conclusion and future work}
\label{sec:conclusion}
In this study, we conducted the first systematic evaluation of neuron coverage criteria on Transformer-based language models, reviewing existing criteria in terms of granularity and algorithmic complexity. Our results show that neuron coverage can capture useful test suite properties such as error exposure, distributional diversity, and input generation guidance, but its effectiveness varies across contexts. The marginal gains from adversarial over benign perturbations indicate limited value for error detection, while most criteria, except order-sensitive variants, perform reliably for distributional and categorical diversity. Moreover, combining multiple criteria provides stronger guidance for input generation than any single one, narrowing the mutation search space and improving the discovery of error-revealing inputs.

\noindent \textbf{Future Work.}

\begin{compactitem}[$\bullet$]
	\item \textbf{Coverage Design for Transformer Architectures.}
	Our findings suggest that future research should move beyond coverage criteria originally designed for convolutional neural networks and instead develop criteria specifically tailored to Transformer architectures. In particular, incorporating intermediate variables intrinsic to self-attention mechanisms, such as the query, key, and value representations, may help better capture error-related signals and improve the ability to distinguish erroneous perturbations from noise-like ones. In addition, designing coverage criteria that are robust to variations in input ordering remains an important challenge. Addressing these issues would further advance coverage-guided testing toward a practical and effective tool for improving the reliability of Transformer-based NLP systems.
	
	\item  \textbf{Efficient Evaluation Configurations.}
	Another promising direction is to investigate alternative evaluation configurations, such as selecting specific tokens or layers, in order to better balance coverage effectiveness and computational cost. This problem naturally involves optimization over subsets of tokens or layers, where the number of possible configurations grows exponentially. Systematically studying how different token- or layer-selection strategies affect coverage sensitivity, diversity characterization, and guidance quality may provide insights into more efficient coverage estimation. Such efforts could ultimately lead to more scalable and targeted coverage-guided testing frameworks for large-scale Transformer models.

    \item \textbf{Optimization of Initialization Data and Test Suite Scale.} 
    The calculation of neuron coverage criteria necessitates a specified test suite size for evaluation and, in many cases, a baseline dataset for initializing neuron states (e.g., determining activation bounds). Currently, these configurations are largely determined empirically, such as the 1,000 and 5,000 sample sizes adopted in this study, representing a subjective trade-off between statistical stability and computational overhead. However, given the immense volume of training data and the prohibitive inference costs of modern Large Language Models (LLMs), relying on arbitrary or "the larger, the better" strategies is untenable. A critical avenue for future work is to investigate methods for identifying the most representative or "critical" subset of data for robust initialization, and to establish scientifically grounded guidelines for determining the optimal test suite size. This would ensure that coverage assessments are both statistically reliable and computationally feasible in the era of large-scale models.

	\item  \textbf{Extending Coverage-Guided Testing to Other Attack Paradigms.}
	Beyond classical adversarial perturbations, recent attack paradigms such as prompt injection have emerged as critical security concerns for large language models. These attacks involve deliberate, semantically meaningful manipulations of input prompts and differ fundamentally from noise-like perturbations considered in this study. Exploring how coverage criteria can be adapted to characterize and detect such attacks, for example by analyzing internal state shifts induced by injected instructions, represents an important and challenging direction for future work. Addressing this problem will likely require new threat models, mutation strategies, and evaluation frameworks, enabling coverage-guided testing to extend beyond robustness testing toward broader security and safety assessments.
\end{compactitem}

%% file: ref.bib
@article{yan2024revisiting,
  title={Revisiting deep neural network test coverage from the test effectiveness perspective},
  author={Yan, Ming and Chen, Junjie and Cao, Xuejie and Wu, Zhuo and Kang, Yuning and Wang, Zan},
  journal={Journal of Software: Evolution and Process},
  volume={36},
  number={4},
  pages={e2561},
  year={2024},
  publisher={Wiley Online Library}
}

@article{geirhos2018imagenet,
  title={ImageNet-trained CNNs are biased towards texture; increasing shape bias improves accuracy and robustness},
  author={Geirhos, Robert and Rubisch, Patricia and Michaelis, Claudio and Bethge, Matthias and Wichmann, Felix A and Brendel, Wieland},
  journal={arXiv preprint arXiv:1811.12231},
  year={2018}
}

@article{brendel2019approximating,
  title={Approximating cnns with bag-of-local-features models works surprisingly well on imagenet},
  author={Brendel, Wieland and Bethge, Matthias},
  journal={arXiv preprint arXiv:1904.00760},
  year={2019}
}

@inproceedings{wang2024survey,
  title={A Survey on Test Input Selection and Prioritization for Deep Neural Networks},
  author={Wang, Shengrong and Li, Dongcheng and Li, Hui and Zhao, Man and Wong, W Eric},
  booktitle={2024 10th International Symposium on System Security, Safety, and Reliability (ISSSR)},
  pages={232--243},
  year={2024},
  organization={IEEE}
}

@misc{OpenAI2023, 
    author = {OpenAI},
    title = {ChatGPT},
    year = {2023}, 
    note = {Retrieved July 26, 2023},
    url = {https://openai.com/chatgpt}
}

@book{ammann2016introduction,
  title={Introduction to software testing},
  author={Ammann, Paul and Offutt, Jeff},
  year={2016},
  publisher={Cambridge University Press}
}

@inproceedings{deepxplore,
  title={Deepxplore: Automated whitebox testing of deep learning systems},
  author={Pei, Kexin and Cao, Yinzhi and Yang, Junfeng and Jana, Suman},
  booktitle={proceedings of the 26th Symposium on Operating Systems Principles},
  pages={1--18},
  year={2017}
}

@article{deepcover,
  title={Testing deep neural networks},
  author={Sun, Youcheng and Huang, Xiaowei and Kroening, Daniel and Sharp, James and Hill, Matthew and Ashmore, Rob},
  journal={arXiv preprint arXiv:1803.04792},
  year={2018}
}

@inproceedings{importance,
	title={Importance-driven deep learning system testing},
	author={Gerasimou, Simos and Eniser, Hasan Ferit and Sen, Alper and Cakan, Alper},
	booktitle={Proceedings of the ACM/IEEE 42nd International Conference on Software Engineering},
	pages={702--713},
	year={2020}
}

@inproceedings{deepgauge,
  title={Deepgauge: Multi-granularity testing criteria for deep learning systems},
  author={Ma, Lei and Juefei-Xu, Felix and Zhang, Fuyuan and Sun, Jiyuan and Xue, Minhui and Li, Bo and Chen, Chunyang and Su, Ting and Li, Li and Liu, Yang and others},
  booktitle={Proceedings of the 33rd ACM/IEEE international conference on automated software engineering},
  pages={120--131},
  year={2018}
}

@inproceedings{deepct,
  title={Deepct: Tomographic combinatorial testing for deep learning systems},
  author={Ma, Lei and Juefei-Xu, Felix and Xue, Minhui and Li, Bo and Li, Li and Liu, Yang and Zhao, Jianjun},
  booktitle={2019 IEEE 26th International Conference on Software Analysis, Evolution and Reengineering (SANER)},
  pages={614--618},
  year={2019},
  organization={IEEE}
}

@inproceedings{deeppath,
  title={DeepPath: Path-driven testing criteria for deep neural networks},
  author={Wang, Dong and Wang, Ziyuan and Fang, Chunrong and Chen, Yanshan and Chen, Zhenyu},
  booktitle={2019 IEEE International Conference On Artificial Intelligence Testing (AITest)},
  pages={119--120},
  year={2019},
  organization={IEEE}
}

@article{deepinspect,
  author       = {Yuchi Tian and
                  Ziyuan Zhong and
                  Vicente Ordonez and
                  Baishakhi Ray},
  title        = {Testing Deep Neural Network based Image Classifiers},
  journal      = {arXiv preprint arXiv:1905.07831},
  year         = {2019}
  
}

@article{deepcruiser,
  title={Deepcruiser: Automated guided testing for stateful deep learning systems},
  author={Du, Xiaoning and Xie, Xiaofei and Li, Yi and Ma, Lei and Zhao, Jianjun and Liu, Yang},
  journal={arXiv preprint arXiv:1812.05339},
  year={2018}
}

@inproceedings{surpriseadequacy,
  title={Guiding deep learning system testing using surprise adequacy},
  author={Kim, Jinhan and Feldt, Robert and Yoo, Shin},
  booktitle={2019 IEEE/ACM 41st International Conference on Software Engineering (ICSE)},
  pages={1039--1049},
  year={2019},
  organization={IEEE}
}

@article{vaswani2017attention,
  title={Attention is all you need},
  author={Vaswani, Ashish and Shazeer, Noam and Parmar, Niki and Uszkoreit, Jakob and Jones, Llion and Gomez, Aidan N and Kaiser, {\L}ukasz and Polosukhin, Illia},
  journal={Advances in neural information processing systems},
  volume={30},
  year={2017}
}

@article{radford2018improving,
  title={Improving language understanding by generative pre-training},
  author={Radford, Alec and Narasimhan, Karthik and Salimans, Tim and Sutskever, Ilya and others},
  year={2018},
  publisher={OpenAI}
}

@article{yang2019xlnet,
  title={Xlnet: Generalized autoregressive pretraining for language understanding},
  author={Yang, Zhilin and Dai, Zihang and Yang, Yiming and Carbonell, Jaime and Salakhutdinov, Russ R and Le, Quoc V},
  journal={Advances in neural information processing systems},
  volume={32},
  year={2019}
}

@article{raffel2020exploring,
  title={Exploring the limits of transfer learning with a unified text-to-text transformer},
  author={Raffel, Colin and Shazeer, Noam and Roberts, Adam and Lee, Katherine and Narang, Sharan and Matena, Michael and Zhou, Yanqi and Li, Wei and Liu, Peter J},
  journal={The Journal of Machine Learning Research},
  volume={21},
  number={1},
  pages={5485--5551},
  year={2020},
  publisher={JMLRORG}
}

@article{sanh2019distilbert,
  title={DistilBERT, a distilled version of BERT: smaller, faster, cheaper and lighter},
  author={Sanh, Victor and Debut, Lysandre and Chaumond, Julien and Wolf, Thomas},
  journal={arXiv preprint arXiv:1910.01108},
  year={2019}
}

@article{jiao2019tinybert,
  title={Tinybert: Distilling bert for natural language understanding},
  author={Jiao, Xiaoqi and Yin, Yichun and Shang, Lifeng and Jiang, Xin and Chen, Xiao and Li, Linlin and Wang, Fang and Liu, Qun},
  journal={arXiv preprint arXiv:1909.10351},
  year={2019}
}

@inproceedings{nlc,
  title={Revisiting neuron coverage for dnn testing: A layer-wise and distribution-aware criterion},
  author={Yuan, Yuanyuan and Pang, Qi and Wang, Shuai},
  booktitle={2023 IEEE/ACM 45th International Conference on Software Engineering (ICSE)},
  pages={1200--1212},
  year={2023},
  organization={IEEE}
}

@article{npc,
  title={Npc: Neuron path coverage via characterizing decision logic of deep neural networks},
  author={Xie, Xiaofei and Li, Tianlin and Wang, Jian and Ma, Lei and Guo, Qing and Juefei-Xu, Felix and Liu, Yang},
  journal={ACM Transactions on Software Engineering and Methodology (TOSEM)},
  volume={31},
  number={3},
  pages={1--27},
  year={2022},
  publisher={ACM New York, NY}
}

@inproceedings{textfooler,
  title={Is bert really robust? a strong baseline for natural language attack on text classification and entailment},
  author={Jin, Di and Jin, Zhijing and Zhou, Joey Tianyi and Szolovits, Peter},
  booktitle={Proceedings of the AAAI conference on artificial intelligence},
  volume={34},
  number={05},
  pages={8018--8025},
  year={2020}
}

@article{devlin2018bert,
  title={Bert: Pre-training of deep bidirectional transformers for language understanding},
  author={Devlin, Jacob and Chang, Ming-Wei and Lee, Kenton and Toutanova, Kristina},
  journal={arXiv preprint arXiv:1810.04805},
  year={2018}
}

@article{lan2019albert,
  title={Albert: A lite bert for self-supervised learning of language representations},
  author={Lan, Zhenzhong and Chen, Mingda and Goodman, Sebastian and Gimpel, Kevin and Sharma, Piyush and Soricut, Radu},
  journal={arXiv preprint arXiv:1909.11942},
  year={2019}
}

@article{liu2019roberta,
  title={Roberta: A robustly optimized bert pretraining approach},
  author={Liu, Yinhan and Ott, Myle and Goyal, Naman and Du, Jingfei and Joshi, Mandar and Chen, Danqi and Levy, Omer and Lewis, Mike and Zettlemoyer, Luke and Stoyanov, Veselin},
  journal={arXiv preprint arXiv:1907.11692},
  year={2019}
}

@article{bowman2015large,
  title={A large annotated corpus for learning natural language inference},
  author={Bowman, Samuel R and Angeli, Gabor and Potts, Christopher and Manning, Christopher D},
  journal={arXiv preprint arXiv:1508.05326},
  year={2015}
}

@inproceedings{harel2020neuron,
  title={Is neuron coverage a meaningful measure for testing deep neural networks?},
  author={Harel-Canada, Fabrice and Wang, Lingxiao and Gulzar, Muhammad Ali and Gu, Quanquan and Kim, Miryung},
  booktitle={Proceedings of the 28th ACM Joint Meeting on European Software Engineering Conference and Symposium on the Foundations of Software Engineering},
  pages={851--862},
  year={2020}
}

@inproceedings{yan2020correlations,
  title={Correlations between deep neural network model coverage criteria and model quality},
  author={Yan, Shenao and Tao, Guanhong and Liu, Xuwei and Zhai, Juan and Ma, Shiqing and Xu, Lei and Zhang, Xiangyu},
  booktitle={Proceedings of the 28th ACM Joint Meeting on European Software Engineering Conference and Symposium on the Foundations of Software Engineering},
  pages={775--787},
  year={2020}
}

@inproceedings{ji2023cc,
  title={Cc: Causality-aware coverage criterion for deep neural networks},
  author={Ji, Zhenlan and Ma, Pingchuan and Yuan, Yuanyuan and Wang, Shuai},
  booktitle={2023 IEEE/ACM 45th International Conference on Software Engineering (ICSE)},
  pages={1788--1800},
  year={2023},
  organization={IEEE}
}

@article{brown2020language,
  title={Language models are few-shot learners},
  author={Brown, Tom and Mann, Benjamin and Ryder, Nick and Subbiah, Melanie and Kaplan, Jared D and Dhariwal, Prafulla and Neelakantan, Arvind and Shyam, Pranav and Sastry, Girish and Askell, Amanda and others},
  journal={Advances in neural information processing systems},
  volume={33},
  pages={1877--1901},
  year={2020}
}

@article{chowdhery2023palm,
  title={Palm: Scaling language modeling with pathways},
  author={Chowdhery, Aakanksha and Narang, Sharan and Devlin, Jacob and Bosma, Maarten and Mishra, Gaurav and Roberts, Adam and Barham, Paul and Chung, Hyung Won and Sutton, Charles and Gehrmann, Sebastian and others},
  journal={Journal of Machine Learning Research},
  volume={24},
  number={240},
  pages={1--113},
  year={2023}
}

@article{touvron2023llama,
  title={Llama: Open and efficient foundation language models},
  author={Touvron, Hugo and Lavril, Thibaut and Izacard, Gautier and Martinet, Xavier and Lachaux, Marie-Anne and Lacroix, Timoth{\'e}e and Rozi{\`e}re, Baptiste and Goyal, Naman and Hambro, Eric and Azhar, Faisal and others},
  journal={arXiv preprint arXiv:2302.13971},
  year={2023}
}

@inproceedings{li2019structural,
  title={Structural coverage criteria for neural networks could be misleading},
  author={Li, Zenan and Ma, Xiaoxing and Xu, Chang and Cao, Chun},
  booktitle={2019 IEEE/ACM 41st International Conference on Software Engineering: New Ideas and Emerging Results (ICSE-NIER)},
  pages={89--92},
  year={2019},
  organization={IEEE}
}

@inproceedings{yang2022revisiting,
  title={Revisiting neuron coverage metrics and quality of deep neural networks},
  author={Yang, Zhou and Shi, Jieke and Asyrofi, Muhammad Hilmi and Lo, David},
  booktitle={2022 IEEE International Conference on Software Analysis, Evolution and Reengineering (SANER)},
  pages={408--419},
  year={2022},
  organization={IEEE}
}

@article{dong2019there,
  title={There is limited correlation between coverage and robustness for deep neural networks},
  author={Dong, Yizhen and Zhang, Peixin and Wang, Jingyi and Liu, Shuang and Sun, Jun and Hao, Jianye and Wang, Xinyu and Wang, Li and Dong, Jin Song and Ting, Dai},
  journal={arXiv preprint arXiv:1911.05904},
  year={2019}
}

@article{badar2020application,
  title={Application of deep learning for retinal image analysis: A review},
  author={Badar, Maryam and Haris, Muhammad and Fatima, Anam},
  journal={Computer Science Review},
  volume={35},
  pages={100203},
  year={2020},
  publisher={Elsevier}
}

@article{lauriola2022introduction,
  title={An introduction to deep learning in natural language processing: Models, techniques, and tools},
  author={Lauriola, Ivano and Lavelli, Alberto and Aiolli, Fabio},
  journal={Neurocomputing},
  volume={470},
  pages={443--456},
  year={2022},
  publisher={Elsevier}
}

@article{nair2020exploring,
  title={Exploring uncertainty measures in deep networks for multiple sclerosis lesion detection and segmentation},
  author={Nair, Tanya and Precup, Doina and Arnold, Douglas L and Arbel, Tal},
  journal={Medical image analysis},
  volume={59},
  pages={101557},
  year={2020},
  publisher={Elsevier}
}

@article{loftus2022uncertainty,
  title={Uncertainty-aware deep learning in healthcare: a scoping review},
  author={Loftus, Tyler J and Shickel, Benjamin and Ruppert, Matthew M and Balch, Jeremy A and Ozrazgat-Baslanti, Tezcan and Tighe, Patrick J and Efron, Philip A and Hogan, William R and Rashidi, Parisa and Upchurch Jr, Gilbert R and others},
  journal={PLOS digital health},
  volume={1},
  number={8},
  pages={e0000085},
  year={2022},
  publisher={Public Library of Science San Francisco, CA USA}
}

@article{egger2022medical,
  title={Medical deep learning—A systematic meta-review},
  author={Egger, Jan and Gsaxner, Christina and Pepe, Antonio and Pomykala, Kelsey L and Jonske, Frederic and Kurz, Manuel and Li, Jianning and Kleesiek, Jens},
  journal={Computer methods and programs in biomedicine},
  volume={221},
  pages={106874},
  year={2022},
  publisher={Elsevier}
}

@article{loquercio2020general,
  title={A general framework for uncertainty estimation in deep learning},
  author={Loquercio, Antonio and Segu, Mattia and Scaramuzza, Davide},
  journal={IEEE Robotics and Automation Letters},
  volume={5},
  number={2},
  pages={3153--3160},
  year={2020},
  publisher={IEEE}
}

@inproceedings{choi2019gaussian,
  title={Gaussian yolov3: An accurate and fast object detector using localization uncertainty for autonomous driving},
  author={Choi, Jiwoong and Chun, Dayoung and Kim, Hyun and Lee, Hyuk-Jae},
  booktitle={Proceedings of the IEEE/CVF International conference on computer vision},
  pages={502--511},
  year={2019}
}

@article{radford2019language,
  title={Language models are unsupervised multitask learners},
  author={Radford, Alec and Wu, Jeffrey and Child, Rewon and Luan, David and Amodei, Dario and Sutskever, Ilya and others},
  journal={OpenAI blog},
  volume={1},
  number={8},
  pages={9},
  year={2019}
}

@article{ouyang2022training,
  title={Training language models to follow instructions with human feedback},
  author={Ouyang, Long and Wu, Jeffrey and Jiang, Xu and Almeida, Diogo and Wainwright, Carroll and Mishkin, Pamela and Zhang, Chong and Agarwal, Sandhini and Slama, Katarina and Ray, Alex and others},
  journal={Advances in Neural Information Processing Systems},
  volume={35},
  pages={27730--27744},
  year={2022}
}

@article{bubeck2023sparks,
  title={Sparks of artificial general intelligence: Early experiments with gpt-4},
  author={Bubeck, S{\'e}bastien and Chandrasekaran, Varun and Eldan, Ronen and Gehrke, Johannes and Horvitz, Eric and Kamar, Ece and Lee, Peter and Lee, Yin Tat and Li, Yuanzhi and Lundberg, Scott and others},
  journal={arXiv preprint arXiv:2303.12712},
  year={2023}
}

@inproceedings{du2019deepstellar,
	title={Deepstellar: Model-based quantitative analysis of stateful deep learning systems},
	author={Du, Xiaoning and Xie, Xiaofei and Li, Yi and Ma, Lei and Liu, Yang and Zhao, Jianjun},
	booktitle={Proceedings of the 2019 27th ACM joint meeting on European software engineering conference and symposium on the foundations of software engineering},
	pages={477--487},
	year={2019}
}

@inproceedings{zheng2019testing,
	title={Testing untestable neural machine translation: An industrial case},
	author={Zheng, Wujie and Wang, Wenyu and Liu, Dian and Zhang, Changrong and Zeng, Qinsong and Deng, Yuetang and Yang, Wei and He, Pinjia and Xie, Tao},
	booktitle={2019 IEEE/ACM 41st International Conference on Software Engineering: Companion Proceedings (ICSE-Companion)},
	pages={314--315},
	year={2019},
	organization={IEEE}
}
